\documentclass[journal]{IEEEtran}
\ifCLASSINFOpdf
\else
\fi

\usepackage{cite}
\usepackage{amsmath,amssymb,amsfonts}
\usepackage{algorithmic}
\usepackage{graphicx}
\usepackage{textcomp}
\usepackage{xcolor}
\usepackage{booktabs}
\usepackage{graphicx}
\usepackage{caption}
\usepackage{subcaption}
\usepackage{siunitx, soul}
\usepackage{adjustbox}
\usepackage{booktabs}
\usepackage{multirow}
\usepackage{url} % hyperref works too
\usepackage[colorlinks]{hyperref} 
\hypersetup{citecolor=green, linkcolor=black, urlcolor=blue}
\usepackage[acronym,toc,nomain]{glossaries}
\newacronym{avc}{AVC}{Advanced Video Coding}
\newacronym{ai}{AI}{All Intra}
\newacronym{alf}{ALF}{Adaptive Loop Filter}
\newacronym{aom}{AOM}{Alliance for Open Media}
\newacronym{amvp}{AMVP}{Advanced Motion Vector Prediction}
\newacronym{avx}{AVX}{Advanced Vector eXtensions}

\newacronym{bt}{BT}{Binary Tree}
\newacronym{bh}{BTH}{Binary Tree Horizontal}
\newacronym{bv}{BTV}{Binary Tree Vertical}
\newacronym{bdr}{BD-BR}{Bj\o ntegaard Delta BitRate}
\newacronym{bdpsnr}{BD-PSNR}{Bj\o ntegaard Delta PSNR}
\newacronym{bdof}{BDOF}{Bi-Directional Optical Flow}
\newacronym{fvc}{FVC}{Future Video Coding}

\newacronym{ctu}{CTU}{Coding Tree Unit}
\newacronym{cu}{CU}{Coding Unit}
\newacronym{cnn}{CNN}{Convolution Neural Network}
\newacronym{ctc}{CTC}{Common Test Conditions}
\newacronym{crc}{CRC}{Complexity Reduction Configuration}
\newacronym{cpu}{CPU}{Central Processing Unit}
\newacronym{cabac}{CABAC}{Context Adaptive Binary Arithmetic Coding}
\newacronym{CTC}{CTC}{Common Test Conditions}
\newacronym{ccitt}{CCITT}{International Telegraph and Telephone Consultative Committee}
\newacronym{cb}{CB}{Coding Block}
\newacronym{ciip}{CIIP}{Combined Intra/Inter Prediction}
\newacronym{cc-alf}{CC-ALF}{Cross Component ALF}
\newacronym{cclm}{CCLM}{Cross Component Linear Model}

\newacronym{dl}{DL}{Deep Learning}
\newacronym{DCT}{DCT}{Discrete Cosine Transform}
\newacronym{DST}{DST}{Discrete Sine Transform}
\newacronym{dmvr}{DMVR}{Decoder-side Motion Vector Refinement}
\newacronym{dpb}{DPB}{Decoded Picture Buffer}
\newacronym{dbf}{DBF}{Deblocking Filter}

\newacronym{fhd}{FHD}{Full High Definition}
\newacronym{fps}{fps}{Frames Per Second}

\newacronym{gop}{GOP}{Group of Pictures}
\newacronym{gpu}{GPU}{Graphical Processing Unit}

\newacronym{hevc}{HEVC}{High Efficiency Video Coding}
\newacronym{hdr}{HDR}{High Dynamic Range}
\newacronym{sdr}{SDR}{Standard Dynamic Range}
\newacronym{hi}{HI}{Horizontal Inconsistency}
\newacronym{isp}{ISP}{Intra Sub-Partitionning}
\newacronym{hm}{HM}{HEVC Model}
\newacronym{hd}{HD}{High Definition}
\newacronym{hfr}{HFR}{High Frame Rate}
\newacronym{hvs}{HVS}{Human Visual System}
\newacronym{svr}{SVR}{Support Vector Machine}

\newacronym{iso}{ISO}{International Organization for Standardization}
\newacronym{iec}{IEC}{International Electrotechnical Commission}
\newacronym{itu}{ITU}{International Telecommunication Union}
\newacronym{ict}{ICT}{Inter-Component Transform}

\newacronym{jvet}{JVET}{Joint Video Exploration Team}
\newacronym{jem}{JEM}{Joint Exploration Model}

\newacronym{qt}{QT}{Quad Tree}
\newacronym{qp}{QP}{Quantization Parameter}
\newacronym{qtbt}{QTBT}{Quad Tree Binary Tree}

\newacronym{lmcs}{LMCS}{Luma Mapping with Chroma Scaling}
\newacronym{ld}{LD}{Low-Delay}
\newacronym{ldp}{LDP}{Low-Delay P}
\newacronym{ldb}{LDB}{Low-Delay B}
\newacronym{lfnst}{LFNST}{Low-Frequency Non-Separable Transform}

\newacronym{mpm}{MPM}{Most Probable Mode}
\newacronym{mv}{MV}{Motion Vector}
\newacronym{ml}{ML}{Machine Learning}
\newacronym{mi}{MI}{Mutual Information}
\newacronym{mdf}{MDF}{Motion Divergence Field}
\newacronym{mc}{MC}{Motion Compensation}
\newacronym{me}{ME}{Motion Estimation}
\newacronym{mpeg}{MPEG}{Moving Picture Experts Group}
\newacronym{mtt}{MTT}{Multi-Type Tree}
\newacronym{MOS}{MOS}{Mean Opinion Score}
\newacronym{MTS}{MTS}{Multiple Transform Selection}
\newacronym{mip}{MIP}{Matrix-based Intra Prediction}
\newacronym{mimd}{MIMD}{Multiple Instructions on Multiple Data}
\newacronym{mse}{MSE}{Mean Squared Error}
\newacronym{pdpc}{PDPC}{Position Dependent intra Prediction Combination}
\newacronym{mrl}{MRL}{Multiple Reference Line}

\newacronym{pdf}{Pdf}{Probability Density Function}
\newacronym{pu}{PU}{Prediction Unit}
\newacronym{psnr}{PSNR}{Peak Signal to Noise Ratio}
\newacronym{pmd}{PMD}{Pyramid Motion Divergence}
\newacronym{poc}{POC}{Picture Order Count}
\newacronym{pps}{PPS}{Picture Parameter Set}

\newacronym{rd}{RD}{Rate Distorsion}
\newacronym{rdo}{RDO}{Rate Distorsion Optimization}
\newacronym{ra}{RA}{Random Access}
\newacronym{rgb}{RGB}{Red Blue Green}
\newacronym{rf}{RF}{Random Forest}
\newacronym{rs}{RS}{Rectangular Slice}

\newacronym{satd}{SATD}{Sum of Absolute Transform Differences}
\newacronym{svm}{SVM}{Support Vector Machines}
\newacronym{si}{SI}{Spatial Information}
\newacronym{scc}{SCC}{Screen Coding Content}
\newacronym{simd}{SIMD}{Single Instruction on Multiple Data}
\newacronym{shvc}{SHVC}{Scalable High Efficiency Video Coding}
\newacronym{sao}{SAO}{Sample Adaptive Offset}
\newacronym{SPS}{SPS}{Sequence Parameter Set}
\newacronym{sisd}{SISD}{Single Instruction on Single Data}
\newacronym{ssim}{SSIM}{Structural SIMilarity}
\newacronym{sbtmvp}{SbTMVP}{Subblock-based Temporal Motion Vector Prediction}
\newacronym{samviq}{SAMVIQ}{Subjective Assessment Methodology for Video Quality}
\newacronym{sse}{SSE}{Streaming SIMD Extensions}

\newacronym{tu}{TU}{Transform Unit}
\newacronym{ti}{TI}{Temporal Information}
\newacronym{tt}{TT}{Ternary Tree}
\newacronym{tth}{TTH}{Ternary Tree Horizontal}
\newacronym{ttv}{TTV}{Ternary Tree Vertical}
\newacronym{trs}{TRS}{Tile and Rectangular Slice}
\newacronym{tl}{TL}{Temporal Layer}
\newacronym{tmvp}{TMVP}{Temporal Motion Vector Prediction}

\newacronym{uhd}{UHD}{Ultra High Definition}

\newacronym{vvc}{VVC}{Versatile Video Coding}
\newacronym{vi}{VI}{Vertical Inconsistency}
\newacronym{vceg}{VCEG}{Video Coding Experts Group}
\newacronym{vtm}{VTM}{VVC Test Model}
\newacronym{vmaf}{VMAF}{Video Multi-Method Assessment Fusion}

\newacronym{wpp}{WPP}{Wavefront Parallel Processing}
\newacronym{waip}{WAIP}{Wide Angular Intra Prediction}

\let\oldtabular=\tabular
\def\tabular{\small\oldtabular}

\newcommand{\Figure}[1] {Fig.~#1}
\newcommand{\Table}[1] {TABLE~#1}
\usepackage{bbding} 

\newcommand{\adcomment}[1]{{\color{black}{#1}}}

\begin{document}
%\UseRawInputEncoding
%\tolerance 3000

\title{Performance and Coding Complexity of the Versatile Video Coding}
\title{Video Quality Assessment and Coding Complexity of the Versatile Video Coding Standard}

\author{
        Thomas~Amestoy, %~\IEEEmembership{Member,~IEEE,}
        Naty~Sidaty, %~\IEEEmembership{Member,~IEEE,}
        Wassim~Hamidouche, %~\IEEEmembership{Member,~IEEE,}
        Pierrick~Philippe and %~\IEEEmembership{Senior,~IEEE}
        Daniel~Menard   %~\IEEEmembership{Senior,~IEEE}
    %    \thanks{T. Amestoy is with Thales~SIX~GTS~France, HTE/STR/MMP Gennevilliers, France, (Email: Thomas.Amestoy@thalesgroup.com)}
\thanks{N. Sidaty, T. Amestoy, W. Hamidouche and D. Menard were with with Univ. Rennes, INSA Rennes, CNRS, IETR - UMR 6164, Rennes, France, (e-mail: \href{mailto:thomas.amestoy@insa-rennes.fr}{thomas.amestoy@insa-rennes.fr} and \href{mailto:wassim.hamidouche@insa-rennes.fr}{wassim.hamidouche@insa-rennes.fr})}% <-this % stops a space
\thanks{P. Philippe was with Orange Lab, 1219 Avenue des Champs Blancs, 35510 Cesson-S\'evign\'e, France, e-mails: (\href{mailto:pierrick.philippe@orange.com}{pierrick.philippe@orange.com})}
}
	\maketitle
	%\tableofcontents
	
\begin{abstract}
In recent years, the proliferation of multimedia applications and formats, such as IPTV, Virtual Reality (VR, 360$^{\circ}$), and point cloud videos, has presented new challenges to the video compression research community. Simultaneously, there has been a growing demand from users for higher resolutions and improved visual quality. To further enhance coding efficiency, a new video coding standard, \gls{vvc}, was introduced in July 2020. This paper conducts a comprehensive analysis of coding performance and complexity for the latest \gls{vvc} standard in comparison to its predecessor, \gls{hevc}. The study employs a diverse set of test sequences, covering both \gls{hd} and \gls{uhd} resolutions, and spans a wide range of bit-rates. These sequences are encoded using the reference software encoders of  \gls{hevc} (\acrshort{hm}) and \gls{vvc} (\acrshort{vtm}). The results consistently demonstrate that \gls{vvc} outperforms \gls{hevc}, achieving bit-rate savings of up to 40\% on the subjective quality scale, particularly at realistic bit-rates and quality levels. Objective quality metrics, including \acrshort{psnr}, \acrshort{ssim}, and \acrshort{vmaf}, support these findings, revealing bit-rate savings ranging from 31\% to 40\%, depending on the video content, spatial resolution, and the selected quality metric. However, these improvements in coding efficiency come at the cost of significantly increased computational complexity. On average, our results indicate that the \gls{vvc} decoding process is 1.5 times more complex, while the encoding process becomes at least eight times more complex than that of the \gls{hevc} reference encoder. Our simultaneous profiling of the two standards sheds light on the primary evolutionary differences between them and highlights the specific stages responsible for the observed increase in complexity.
\end{abstract}


	% \fi
	\IEEEpeerreviewmaketitle
	
	\glsresetall

\section{Introduction}\label{section:introduction}
	
	\ifCLASSOPTIONcaptionsoff
	\newpage
	\fi

\IEEEPARstart{O}{ver}  the last decade, there has been a significant surge in multimedia services and video applications, driven by remarkable advancements in digital technologies. Emerging video applications and image representations offer a more immersive and natural viewing experience. However, these new services demand higher quality and resolution, such as 4K and 8K, to meet the quality of service expectations of end users. The widely recognized video coding standard, \gls{hevc}/H.265, has gradually gained adoption in numerous application systems. It offers up to a 50\% reduction in bit-rate while maintaining equal subjective quality compared to its predecessor, \gls{avc}/H.264 standard~\cite{sullivan2012overview, 6317156, 7254155}. In 2015, new coding tools were developed under the Joint Exploration Model (JEM) software, aiming to achieve substantial bit-rate savings compared to the \gls{hevc} standard, and promising results were obtained~\cite{8296832, Alshina}. However, these quality improvements have introduced a significant increase in complexity~\cite{8296832}.

To address these challenges, the Joint Video Experts Team (JVET) issued a call for proposals in 2017 to develop a new video coding standard, initially known as Beyond the \gls{hevc} or \gls{fvc}, now referred to as \gls{vvc}/H.266~\cite{8861138, 9689950}. \gls{vvc}  is poised to facilitate the deployment of higher-quality video services and emerging applications, including \gls{hdr}, \gls{hfr}, and 360-degree omnidirectional immersive multimedia. Completed in July 2020, the primary goal of \gls{vvc}/H.266 is to achieve a substantial improvement in compression performance compared to the existing \gls{hevc}/H.265 standard. The aim is to achieve a bit-rate savings of 40\% or more in terms of \gls{psnr} compared to \gls{hevc}, with a reasonable increase in complexity at both the encoder and decoder. The new standard is anticipated to enable the delivery of Ultra High Definition (4K) and 8K services at bit-rates comparable to those used for HDTV, effectively doubling the amount of multimedia content that can be stored on a server or transmitted through a streaming platforms. However, it is essential to note that the computational complexity and energy consumption of recent video compression standards has become a critical concern~\cite{7952338}. Prior research has delved into the complexity aspects of \gls{hevc} encoding/decoding for various resolutions, including \gls{hd} (1920$\times$1080) and \gls{uhd} (4K)~\cite{10.1117/12.954123, 7273890, 6317152, 6698051}. In the case of JEM tools, the coding gains achieved come at the cost of significantly increased complexity, estimated to be ten times higher for both encoder and decoder in Inter coding configuration.

In this paper, we conduct a comprehensive assessment of the emerging \gls{vvc}/H.266 standard, comparing it to the established \gls{hevc}/H.265 standard. This evaluation encompasses both subjective and objective quality analyses. We utilize the \gls{vvc} reference software model (VTM) and compare it with the \gls{hevc} reference software model (HM) in Random Access (RA) coding configuration. Our study encompasses a diverse set of video content, spanning various bit-rates and two spatial resolutions: \gls{hd} and \gls{uhd}~\cite{8954535}. Additionally, this paper offers an in-depth analysis of time and complexity distributions for both the encoding and decoding processes. To elucidate the principal evolution from one reference software to the other, we concurrently present the profiling of VTM-5.0 and HM-16.20. While a previous study~\cite{6317152} assessed the impact of specific new tools on \gls{vvc}/H.266  complexity by disabling them individually and measuring encoding/decoding times, it is crucial to note that many other tools are mandatory in the primary profile and cannot be disabled on the encoder side. Furthermore, disabling a single tool can lead to different coding decisions compared to the reference configuration. The profiling approach employed in our study~\cite{6317152} offers two significant advantages compared to the aforementioned work: it enables us to evaluate the complexity of all encoding/decoding tools comprehensively, and complexity measurements are conducted within the original encoding/decoding processes with all tools enabled.

The rest of this paper are organized as follows. Section~\ref{Versatile Video Coding Description} gives an introduction to the \gls{vvc} standard, including its architecture and the key coding tools integrated into the VTM. Section~\ref{VVC Coding Performance: Subjective Study} outlines our methodology for subjective quality assessment, detailing the test material, experimental environment, and the methodologies employed. The results obtained from our assessments, along with their associated analysis, are presented in Section~\ref{Experimental Results}. Section~\ref{complexity} offers an in-depth examination of complexity distributions for both the encoding and decoding processes. Finally, Section~\ref{sec:con} concludes the paper.
	
%	Deja utilises: \gls{hm} et le \gls{vtm} et la \gls{hd} et le \gls{qp} et le \gls{ra} et le \gls{cpu} et le \gls{qt} et le \gls{mtt} et le \gls{ctu} et le \gls{tu} et le \gls{alf}.

 \section{VVC Tools Description}
 \label{Versatile Video Coding Description}

 As \gls{hevc}, \gls{vvc} has a block-based hybrid coding architecture, combining Inter and Intra predictions with transform coding. Based on the \gls{hevc} standard, \gls{vvc} refined existing technologies but also added novel coding tools. In this section, we briefly describe the main \gls{vvc} coding tools present in the \gls{vtm}-5.0 codec, including frame partitioning, \gls{MTS} , in-loop filtering,  Intra \& Inter predictions and entropy coding. Most of the potential contributions in \gls{vvc} are provided by a set of tools, given in Table~\ref{tablegain}, with their individual loss when these tools are disabled in the \gls{vtm}-5.0. The authors do not intend here to provide an exhaustive description of the \gls{vvc} tools, for detailed description of these coding tools, the reader is referred to~\cite{jvetdoc, 9689950}.
 
 \subsection{Frame Partitioning}
% In addition to the recursive \gls{qt} partitioning used in \gls{hevc}, \gls{vtm}-5.0 integrates a nested recursive \gls{mtt} partitioning, i.e. \gls{bt} and \gls{tt} splits. 
In \gls{hevc}, the \gls{qt} partitioning suffered from several limitations:
 \begin{itemize}
 	\item \glspl{cu} could only be square, with no flexible shape to cover all block characteristics.
 	\item Luma and Chroma components of the encoded sequence have the same \gls{qt} splitting which is rarely optimal for chroma.
 	\item Residual could only be split into \glspl{tu} with square shapes, reducing the potential impact of transforms.
 \end{itemize}
 
\begin{figure}[t]
 	\centering
 	\includegraphics[width=0.9\linewidth]{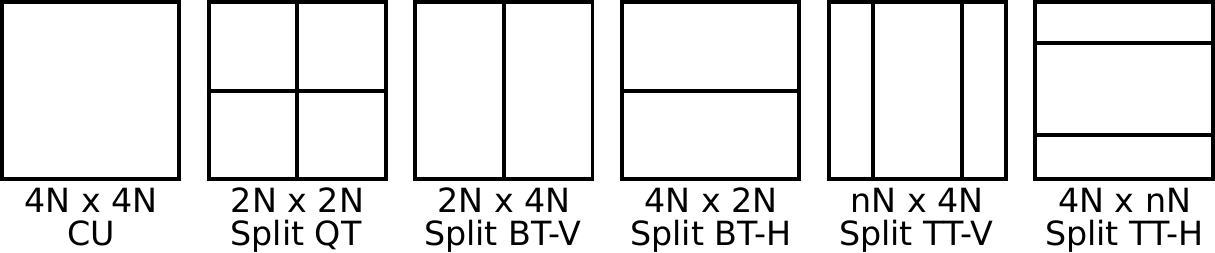}
 	\caption{Available split included in \gls{vtm}-5.0 of a  $4N \times 4N$ \gls{cu}.}
 	\label{fig:split_possible}
 \end{figure}

\begin{figure}[t]
 	\centering
 	\includegraphics[width=0.9\linewidth]{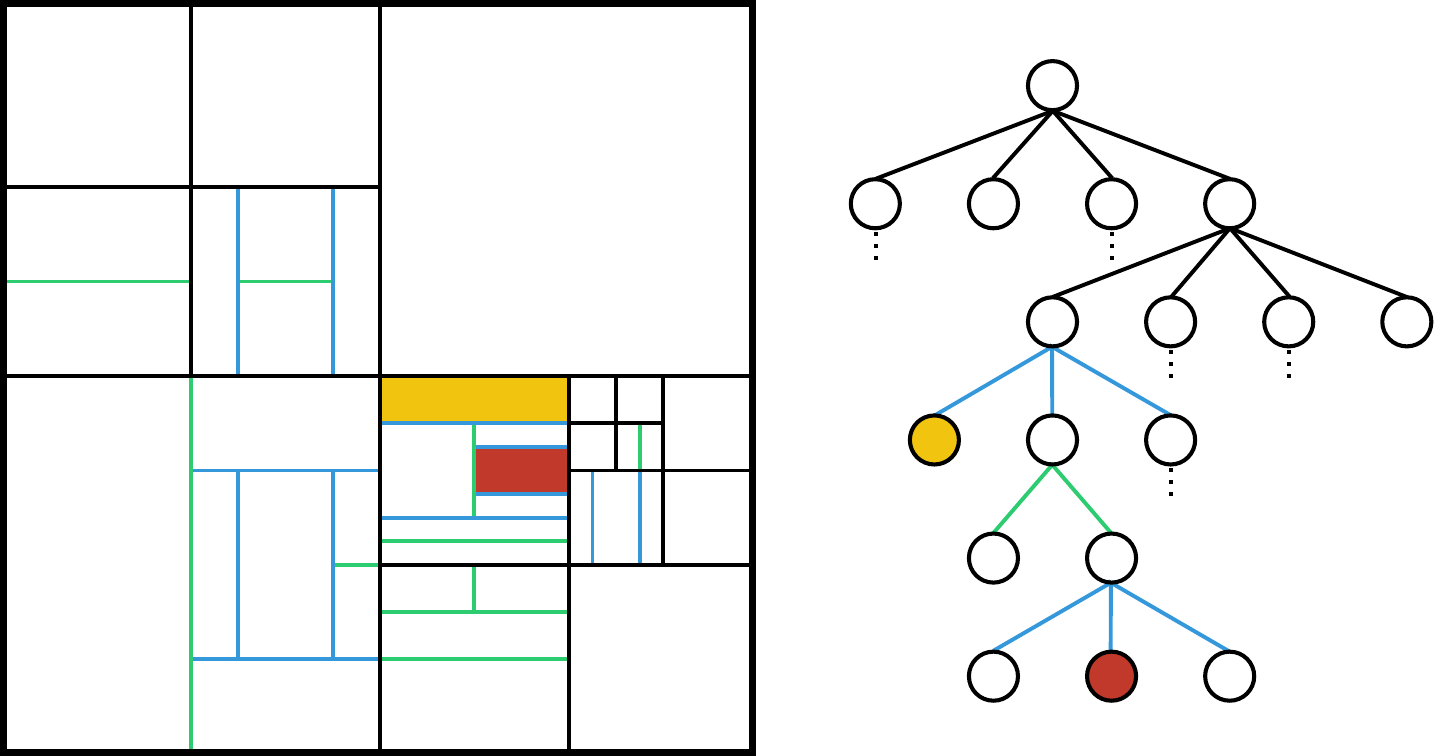}
 	\caption{Example of a \gls{ctu} partitioning with a part of its corresponding tree representation.}
 	\label{fig:split_example}
 \end{figure}

To overcome these limitations, the \gls{vtm} integrates a nested recursive \gls{mtt} partitioning, i.e., \gls{bt} and \gls{tt} splits, in addition to the recursive \gls{qt} partitioning used in \gls{hevc}. 
 \Figure{\ref{fig:split_possible}} illustrates all available split in \gls{vtm} for a  $4N \times 4N$ \gls{cu}. The \gls{bt} partitioning consists of symmetric horizontal splitting (\gls{bt}-H) and symmetric vertical splitting (\gls{bt}-V) while the \gls{tt} partitioning allows horizontal triple-tree splitting (\gls{tt}-H) and vertical triple-tree splitting (\gls{tt}-V) corresponding to split the \gls{cu} in three blocks with the middle blocks size equal to the half of the \gls{cu}.
 %In \gls{ai} configuration file defined in the \gls{ctc}, \gls{bt} and \gls{tt} are available on \glspl{cu} with sizes lower or equal than $64 \times 64$ and $32 \times 32$, respectively. Moreover, the partitioning process includes restrictions to avoid generating the same \gls{cu} with different succession of split. 
 %For example, \gls{bt} partitioning is not allowed on the middle \gls{cu} of a \gls{tt} split in the same direction.
 %The middle \gls{cu} of a \gls{tt} split cannot be split by a \gls{bt} split in the same direction. 
 Once a \gls{bt} or \gls{tt} split has been executed on a \gls{cu}, any further \gls{qt} split is prohibited within its sub-\glspl{cu}. In Figure~\ref{fig:split_example}, the example on the left illustrates the division of a \gls{ctu} into multiple \glspl{cu} after the application of \gls{ctu} partitioning with \gls{qt} and \gls{mtt} splits. The right side of the figure displays a segment of the \gls{ctu} partitioning tree structure, where dark, green, and blue lines denote \gls{qt}, \gls{bt}, and \gls{tt} splits, respectively. The yellow and red backgrounds highlight two instances of the corresponding \glspl{cu} within the \gls{ctu} partitioning and its related tree.
 
%\blue{To overcome the fact that residual could only be split into square \glspl{tu} in \gls{hevc},
To address the limitation of square residual \glspl{tu} observed in \gls{hevc}, the \gls{vtm} has introduced rectangular transforms. This transformation is applied directly to the \glspl{cu}, regardless of their shape, eliminating the need for any additional subdivision of residual blocks~\cite{zhao_joint_2018}. Additionally, the \gls{isp} tool has the capability to further split Intra-predicted blocks either vertically or horizontally into 2 or 4 sub-partitions, depending on the block size. %\Figure{\ref{fig:isp}} illustrates the possible \gls{isp} partitioning configurations depending on the size of the original block. 
\gls{isp} enables a bit-rate saving of 0.41\% in AI coding configuration.  

% \begin{figure}[t]
% 	\centering
% 	\includegraphics[width=1\linewidth]{figures/ISP.pdf}
% 	\caption{Intra sub-partitioning configurations}
% 	\label{fig:isp}
% \end{figure}
%
%  
%With \gls{qtbt} splitting configuration, \glspl{pu}, \glspl{tu} and \glspl{cu} are melt into one. 
The concept of merging \glspl{tu} and \glspl{cu} into a single entity is introduced. Additionally, it's noteworthy that different partition trees can be employed for the Luma and Chroma components in intra-predicted slices within the \gls{vtm}. However, in the case of inter-predicted slices, the partition trees of Luma are utilized for the Chroma components.

 \subsection{Transform Module}
\begin{figure}[t]
 	\includegraphics[width=1.0\linewidth]{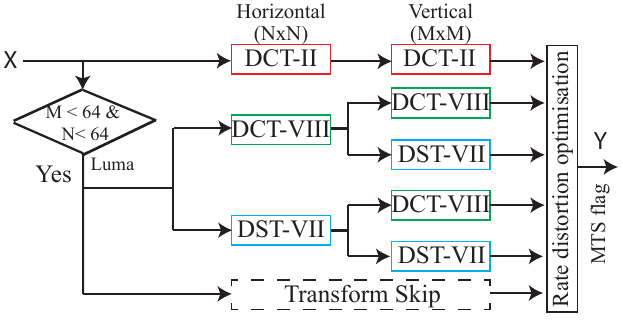}
 	\caption{The concept of 2D separable  transforms selection in \gls{vvc}. $X$ is the input block of residuals, $Y$ is the output transformed block and \gls{MTS} flag is the index of the selected set of transforms.}
 	\label{fig:MTS}
 \end{figure}

  The concept of \gls{MTS} in \gls{vvc} introduces three distinct trigonometric transform types, which are \gls{DCT}-II/VIII and \gls{DST}-VII. In the context of \gls{MTS}, as depicted in \Figure{\ref{fig:MTS}}, these transforms are chosen for Luma blocks smaller than 64, with the selection based on minimizing the rate distortion cost among five different transform sets and the skip configuration. However, for chroma components and Luma blocks of size 64, only \gls{DCT}-II is considered. This utilization of the \gls{MTS} solution results in a significant coding improvement, with gains of approximately 0.84\% in \gls{ai} and 0.33\% in \gls{ra} coding configurations when compared to \gls{hevc}~\cite{JVET-O0013}.

The \textit{sps\_mts\_enabled\_flag}, which is defined in the \gls{SPS}, serves to enable the activation of the \gls{MTS} concept within the encoder. In addition to this flag, two other flags are defined at the \gls{SPS} level, one to signal the use of implicit \gls{MTS} signaling for Intra-coded blocks and another to indicate the use of explicit \gls{MTS} signaling for Inter-coded blocks. In the case of explicit signaling, which is the default mode in the \gls{CTC}, the \textit{tu\_mts\_idx} syntax element is employed to convey information about the selected horizontal and vertical transforms. Finally, this flag is encoded using a Truncated Rice code with a rice parameter $p=0$ and $cMax=4$ (TRp).
 
% \begin{table}
%\centering
%\renewcommand{\arraystretch}{1.145}
%\caption{Signalling of the MTS in the explicit mode}
%\label{table::MTS_synt}  
%\begin{tabular}{|c | c | c |}
%\hline
%\multirow{2}{*}{\it tu\_mts\_idx}   &  \multicolumn{2}{c|}{Transform Direction}  \\
% \cline{2-3}
%    &  Horizontal Transform & Vertical Transform  \\
%    \hline
%    0 & DCT-II & DCT-II \\ 
%    1 & DST-VII & DST-VII \\ 
%    2 & DCT-VIII & DST-VII \\ 
%    3 & DST-VII & DCT-VIII \\ 
%    4 & DCT-VIII & DCT-VIII \\ 
%\hline
%\end{tabular}
%\end{table}
 
The \gls{lfnst} has been incorporated into VTM-5. This technique leverages matrix multiplication to process coefficients from the forward primary transform at the encoder side:
\begin{equation}
\vec{Y} = T \cdot \vec{X},
\end{equation}
where the vector $\vec{X}$ includes the low frequency coefficients of the block rearranged in a vector and the matrix $T$ contains the coefficients transform kernel. The \gls{lfnst} is enabled only when \gls{DCT}-II is used as a primary transform.  

Four sets of two \gls{lfnst} kernels of sizes 16$\times $16 and 64 $\times$64 are applied on 16 coefficients of small block (min (width, height) $<$ 8 ) and 64 coefficients of larger (min (width, height) $>$ 4) blocks, respectively. The \gls{vvc} specification defines four different transform sets selected depending on the Intra prediction mode and each set defines two transform kernels. The used kernel within a set is signaled in the bitstream. To reduce the complexity in number of operations and memory required to store the transform coefficients, the 64$\times$64 inverse transform is reduced to 48$\times$16. Therefore, only 16 basis of the transform kernel is used and the number of input coefficients is reduced to 48 by excluding the bottom right 4$\times$4 block (i.e., include only coefficients of the top-left, top-right and bottom-left 4$\times$4 blocks).% Finally, \gls{lfnst} is enabled only when \gls{DCT}-II is used as primary transform.             

 % %\begin{figure}[t]
 %\begin{minipage}[b]{1.0\linewidth}
 % \centering
 %\centerline{\includegraphics[height= 4cm, width=4.5cm]{fig_PCS/inter}}
 % \vspace{2.0cm}
 %\caption{67 Intra Prediction Modes in VVC.}
 %\label{intra}
 %\end{minipage}
 %\end{figure}

 %\subsection{Block Partitioning}
 %-Highly flexible partitioning \\
 %-Root size 128x128
 %- Multiple splits wich are embedded in multiple tree structures: \\
 %2 sleeting stages: \\
 %1:quad split (block is non split or split in 4 parts)  \\
 %2: binary split (horizontally in two parts or vertically in two parts) or tenary split (horizontally in three parts or vertically in three parts
 \subsection{Intra and Inter Prediction Tools}
 %In \gls{vvc}, 65 intra directional intra prediction modes are used, instead of 33 in \gls{hevc}, in order to capture the arbitrary edge directions presented in a natural video. Moreover, rectangular blocks are used, compared to only square blocks in \gls{hevc}. The Intra prediction modes are coded with a 6 \glspl{mpm} list instead of a list of 3 \glspl{mpm} in \gls{hevc}. In the \gls{vtm}, the results of intra prediction of planar mode are further modified by a \gls{pdpc} method. \gls{pdpc} is an intra prediction method which invokes a combination of the un-filtered boundary reference samples and \gls{hevc} style intra prediction, with filtered boundary reference samples \cite{Bossen, JVET-O0013}. Moreover, the concept of \gls{mrl} intra prediction is introduced in \gls{vvc}. \gls{mrl} uses more reference lines for intra prediction while \gls{hevc} uses only the nearest reference line. The index of the selected reference line is signaled to the decoder in the bitstream. 

 In \gls{vvc}, there are 65 intra directional prediction modes compared to 33 in \gls{hevc}, capturing natural video's diverse edge directions. It employs rectangular blocks instead of just square ones in \gls{hevc}. Intra prediction modes are coded with a 6 \glspl{mpm} list, unlike HEVC's 3 \glspl{mpm} list. The \gls{vtm} refines the planar mode's Intra prediction results using the \gls{pdpc} method, a combination of unfiltered and HEVC-style intra prediction with filtered boundary samples. Additionally, \gls{vvc} introduces \gls{mrl} intra prediction, using \gls{mrl} instead of \gls{hevc}'s nearest reference line, with the index signaled in the bitstream.
 
% \subsection{Inter Prediction Tools}
% For the Inter Prediction, the \gls{vvc} includes several new and refined inter prediction coding tools, including \gls{sbtmvp}, Affine Motion Model (AFF), \gls{bdof}, more advanced \gls{mv} prediction (inherit more information from reference, combine temporal and spatial prediction)~\cite{jvetdoc}. The \gls{dmvr}, also included in \gls{vtm}, reduces the bit-rate induced by Inter prediction by refining the motion vectors at the decoder side.

In the realm of Inter prediction, \gls{vvc} incorporates a range of novel and improved coding tools, such as \gls{sbtmvp}, the Affine Motion Model (AFF), \gls{bdof}, and advanced \gls{mv} prediction that inherits more information from reference, combine temporal and spatial prediction~\cite{jvetdoc}. Additionally, the \gls{dmvr} feature, present in \gls{vtm}, optimizes bit-rate reduction in Inter prediction by refining motion vectors at the decoder.
 
 \subsection{Entropy Coding and In-Loop Filtering}

 In contrast to \gls{hevc}, where transform coefficients of a coding block are coded using non-overlapped Coefficient Groups (CGs), each containing the coefficients of a 4$\times$4 block, \gls{vtm} employs various CGs (1$\times$16, 2$\times$8, 8$\times$2, 2$\times$4, 4$\times$2, and 16$\times$1). Furthermore, significant changes are introduced to the core of the \gls{cabac} engine in \gls{vvc} compared to its design in \gls{hevc}. While the \gls{cabac} engine in \gls{hevc} employs a table-based probability transition process among 64 distinct representative probability states, \gls{vvc} adopts a decode decision model with a 2-state approach and variable probability updating window sizes~\cite{Bross}.

In the realm of in-loop filtering, \gls{vvc} introduces an \gls{alf} alongside the traditional deblocking filter and \gls{sao}. For the luma component, it selects one of 25 filters for each 4$\times$4 block based primarily on local gradient direction and activity. Additionally, \gls{vvc} employs two diamond filter shapes, a 7$\times$7 diamond shape for the luma component and a 5$\times$5 diamond shape for chroma components.

%Compared to \gls{hevc}, where the transform coefficients of a coding block are coded using non-overlapped Coefficient Groups (CGs)  each of them contains the coefficients of a 4x4 block of a coding block, \gls{vtm} uses various CGs (1$\times$16, 2$\times$8, 8$\times$2, 2$\times$4, 4$\times$2 and 16$\times$1). In addition, the core of the \gls{cabac} engine has some important changes in the \gls{vvc} compared to the design in \gls{hevc}. The \gls{cabac} engine in \gls{hevc} uses a table-based probability transition process between 64 different representative probability states. In the \gls{vvc}, a decode decision uses a 2-state model with variable probability updating window sizes~\cite{Bross}. 

% \subsection{ In-Loop Filtering}
 %In  \gls{vvc}, besides deblocking filter and \gls{sao}, an \gls{alf} with block-based filter adaptation is applied. Particularly, for the luma component, one among 25 filters is selected for each 4x4 block. This selection is mainly based on the direction and activity of local gradients. In addition, two diamond filter shapes are used; 7$\times$7 diamond shape for luma component and 5$\times$5 diamond shape for chroma components.

 \begin{table}[tp]
 	\centering
 	\renewcommand{\arraystretch}{1.15}
 	\caption{Performance of the main tools included in the \gls{vtm}-5.0 software~\cite{JVET-O0013}.}
 	\begin{adjustbox}{max width=1\columnwidth}
 	\begin{tabular}{|l|l|c|}
 		\hline
 		Module & Tool description & BD-BR   \\
 		\hline
 		\hline
 		Frame  	 & Triangular partition mode & 0.35\%  \\
 		partitioning &  Chroma separate tree & 0.14\%  \\
 		& Sub-block transform & 0.41\% \\
 		\hline
 		\multirow{2}{*}{Transforms} 	 & Multiple Transform Selection & 0.33\%  \\
 		&  Low-frequency non-separable tran. & 0.79\%  \\
 		\hline
 		& Decoder-side Motion Vector Refinement & 0.82\%\\
		  & Subblock-based temp. merging cand. &	0.43\%\\
 		Inter & Affine motion model	& 2.53\%  \\
 		Prediction & Merge with MVD &0.58\% \\
 		& Bi-directional optical flow & 0.78\% \\
 		& Temporal motion vector predictor & 1.19\% \\
 		\hline
 		Intra     & Multi-reference line prediction  &	0.20\%   \\
 		Prediction & Intra sub-partitioning & 0.13\% \\
 		& Matrix based intra prediction & 0.27\%\\
 		\hline
 		\multirow{1}{*}{In-Loop Filt.}  & Adaptive Loop Filter  &	4.91\%   \\
 		\hline
 		Quantization & Dependent Quantization & 	1.71\%\\
 		%\hline
 		%CABAC & Core CABAC engine &	0.88\%  \\
 		\hline
 	\end{tabular}
	 \end{adjustbox}
 	\label{tablegain} %\vspace{-5mm}
 \end{table}

 %%%%%%%%%%%%%%%%%%%%%%%%%%%%%%%%%%%%%%%%%%%%%%%%%%%%%

 \section{VVC Coding Performance: Quality Enhancements}
 \label{VVC Coding Performance: Subjective Study}
 
 In order to evaluate the coding performance of \gls{vtm} compared to \gls{hevc} \gls{hm}, we have performed  two types of quality evaluations: objective measurements and subjective assessments. These two methods are described bellow. 
 
\subsection {Objective Quality Measurements}

The main goal of an objective quality assessment metric is to produce the highest quality rating
possible close to that provided, on average, by a set of observers during a campaign of subjective tests (see Section \ref{quality assessment}). Several metrics of different categories are used: Mathematical approaches, weighting approaches and by modelling the Human Visual System (HVS).
  \subsubsection {PSNR Metric}
The \gls{psnr} metric is a purely ``signal'' measure. Whatever the nature of the stimuli (image, video), \gls{psnr} is a pixel-to-pixel treatment regardless of the 2D aspect of the image or video. \gls{psnr} is based on the \gls{mse}, to evaluate the error between a degraded image and its reference version. It is defined by~\eqref{eq:psnr}.

\begin{equation}
MSE = \frac{1}{N \, M} \sum\limits_{\substack{i=0}}^{N-1} \sum\limits_{\substack{j=0}}^{M-1} {(Im_{r}(i,j) - Im_{e}(i,j))^2},
\label{eq:psnr}
\end{equation}

where $N$ and $M$ are the horizontal and vertical dimensions of the image, $ Im_ {r} $ is the reference image and $ Im_ {e} $ is the image to be evaluated. 

The \gls{psnr} of a signal of amplitude $A$ is defined by~(\ref {equation6.1}). A \gls{psnr} value greater than $ 40 \, dB $ usually indicates that the image or video to be evaluated is of very good quality.

\begin{equation}
PSNR = 10 \log_{10} (\frac{A^2}{MSE}),    
\label{equation6.1}
\end{equation}

with $A = 2^\gamma-1$ and $\gamma$ is the video bitdepth.

For video sequences, which ordinarily consist of three color components ($Y$, $U$, $V$), we use generally a weighted \gls{psnr} ($wPSNR$) version. The weighted \gls{psnr} is computed by weighting three components $Y$, $U$ and $V$ as bellow

\begin{equation}
    wPSNR = \frac{6 \, PSNR_Y + PSNR_U + PSNR_V}{8}.
    \label{psnr}
\end{equation}

\gls{psnr} is the most used commonly objective metric in all digital domains. It is easy and fast to implement. However, it simply calculates the pixel-to-pixel resemblance without considering the relationships between these pixels. 

 \subsubsection {SSIM Metric}
 
To substantially remedy this disadvantage, Wang {\it et al.}~\cite{1284395} proposed a metric named \gls{ssim} based on the similarities between the structures in the image. Thus, it makes it possible to extract three different criteria, for each ``window'' of the image, judged important for human observers, namely luminance, contrast and structure. It is given by:

\begin{equation}
\begin{split}
& SSIM (Im_{r}, Im_{e})  = \\
&  \frac{(2 \, \mu_{Im_{r}} \mu_{Im_{e}} + C_1)(2 \, \sigma_{{Im_{}}{Im_{e}}} + C_2)}{(\mu_{Im_{r}}^2 + \mu_{Im_{e}}^2 + C_1)(2 \, \sigma_{Im_{r}}^2 + \sigma_{Im_{e}}^2 + C_2)},
\label{equation6.2}
\end{split}
\end{equation}

where $\mu_{Im_{r}}$ is the average neighborhood luminance, $Im_{r}$,  $\sigma_{Im_{r}}$ is the standard deviation of the neighborhood, $Im_{r}$,  $\sigma_{{Im_{r}}{Im_{e}}}$ is the covariance between  $Im_{r}$ and $Im_{e}$. $C_1$ and $C_2$ are two constants that depend on the dynamics of the image.

The values of this metric are then obtained for all calculation windows. They are between 0 and 1: a value close to 1 indicates that both images (videos) have a maximum fidelity while values close to 0 indicate a too degraded image (video). Moreover, \gls{ssim} has been recently exploited~\cite{zhou_ssim-based_2019} in the \gls{hevc} rate-distortion optimization stage to enhance the perceived video quality under a rate control process.  

 \subsubsection {VMAF Metric} 
 \gls{vmaf} is a perceptual video quality assessment algorithm developed by Netflix with collaboration of Laboratory of Image and Video Engineering (LIVE) \cite{VMAF}. \gls{vmaf} is a full-reference, perceptual video quality metric focused on quality degradation due to compression and rescaling. It was specifically formulated to approximate human perception of video quality and then to correlate strongly with subjective \gls{MOS} scores. Using machine learning techniques, a large sample of \gls{MOS} scores were used as ground truth to train a quality estimation model. \gls{vmaf} estimates the perceived quality score by computing scores from multiple quality assessment algorithms and fusing them using a \gls{svm}. \gls{vmaf} has been shown to outperform other image and video quality metrics, such as \gls{ssim} and \gls{psnr} on several datasets in terms of prediction accuracy, when compared to subjective ratings. 
 
  \subsubsection {Bj{\o}ntegaard Metric} 
The {Bj{\o}ntegaard measure is become on of the most popular metric for video coding performance in the recent years. This metric is computed as a difference, in bit-rate, based on interpolating curves from the tested data points \cite{bjontegaard2001calculation}. The comparison used the Bj{\o}ntegaard-Delta bit-rate (BD-BR) measurement method, in which negative values tell how much lower the bit rate is reduced, and positive values tell how much the bit-rate is increased for the same \gls{psnr} quality~\cite{bjontegaard2008}.

\subsection {Subjective Quality  Assessments} 
\label{quality assessment}
Unlike signal-based and HVS-based approaches for objective quality measurements, subjective quality assessment is the process of employing human viewers for grading video quality based on individual perception \cite{7254155}. The global environment, viewing conditions and the implementation process for subjective quality
assessments are specified in various ITU recommendations. Two specific methodologies are widely used in the video quality assessments: ITU-T Rec. P.910 \cite{rec1} for multimedia applications and ITU-R Rec. BT.500 \cite{rec2}, for television pictures.
 
 \subsubsection {Experimental environment}
 The subjective study has been conducted in the INSA/IETR PAVIM Lab, which is a platform for video quality monitoring actively involved in the emerging video contents. This platform includes a psycho-visual testing room, complying with the ITU-R BT.500-13 Recommendation. A display screen Ultra High Definition (4K) of 55 inches Loewe Bild 7.55 was used to visualise the video sequences. 44 observers, 30 men and 14 women aged from 20 to 55 years, have participated in this experiment. All the subjects were screened for color blindness and visual acuity using Ishihara and Snellen charts, respectively, and have a visual acuity of 10/10 in both eyes with or without correction. Finally, all participants have been gratified. Viewers are placed at distances of 1.5 and 3 times the height of the screen for \gls{uhd} and \gls{hd} resolutions, respectively. In this section, we provide information regarding the used video sequences, test material, test settings, and evaluation procedures.
 
 \subsubsection {Test Video Sequences}\label{subsec:test_video}
 In this experiment, a set of high resolution video sequences, from various categories (music, sport, gaming, etc.) has been selected from several datasets (Huawei, SVT, b$<>$com) as well as 4EVER\footnote{For Enhanced Video ExpeRience 2 project, www.4ever-2.com} database. The target resolutions for this test are \gls{hd} (i.e. 1920$\times$1080) and \gls{uhd} (3840$\times$2160), in a \gls{sdr}. Firstly, 13 videos sequences in \gls{uhd} (4K) resolution have been selected and down-sampled using the Scalable SHVC down sampling filters. The choice of these video is mainly based on  the video encoding complexity in terms of colour, movement, texture and homogeneous content.

 \begin{table}[t]
 	\centering
 	\renewcommand{\arraystretch}{1.15}
 	\caption{Configuration parameters for \gls{hm}-16.20 and \gls{vtm} main profiles, in \gls{ra} configuration. }
 	\label{tb:configurations}
 	\begin{adjustbox}{max width=1\columnwidth}
 		\begin{tabular}{lll}
 			\toprule\textbf{Parameter}            &   \textbf{\gls{hm}-16.20 main}    & \textbf{\gls{vtm}-5.0 main} \\\midrule
 			\textbf{\gls{ctu} size}    &  64     &  128  \\
 			\textbf{\gls{qt} max depth}   &  4      &  4    \\
 			\textbf{\gls{mtt} max depth}  &  0      &  3    \\
 			\textbf{Transform Types}      &  DCT-II, DST-VII &  DCT-II, DST-VII,  DCT-VIII \\
 			\textbf{Max \gls{tu} size}    &  32     & 64    \\
 			\textbf{Loop Filters}         &  DBF, SAO     & DBF, SAO, ALF \\
 			\textbf{Search Type}          &  TZ  &  TZ   \\
 			\textbf{Search Range}         &  384 &  384 \\
 			\textbf{N. Ref. Pictures}     &  5   &  5 \\
 			\textbf{Entropy Coding}       &  CABAC    &  CABAC \\
 			\textbf{Internal Bit Depth }  &  10  &  10 \\\bottomrule
 		\end{tabular}
 	\end{adjustbox}
 \end{table}
 
% All encodings are carried out with the main profiles of the \gls{hm}-16.20 and \gls{vtm}-5.0. The profiles configuration parameters are summarized in \Table{\ref{tb:configurations}}.
% As detailed in Section~??, the parameters that differ from \gls{hm}-16.20 to \gls{vtm}-5.0 include \gls{ctu} size, allowed partition modes, Max \gls{tu} size and enabled loop filters.  

All these videos were encoded using both \gls{hevc} (\gls{hm}-16.20) and \gls{vvc} (\gls{vtm}) main profiles, at different bit-rates, in \gls{hd} and \gls{uhd} format in \gls{ra} configurations.  
The main profiles configuration parameters are summarized in\Table{\ref{tb:configurations}}.
 This first selection is done in order to retain the sequences representing a good balance of content variety and video coding artifacts.
 After this initial selection, only seven scenes have been retained for the experiment , as given in Table~\ref{table1}. In total 168 video sequences are used in this study: 7 scenes $\times$ 5 bit-rates $\times$ 2 codecs $\times$ 2 resolutions $+$  28 original scenes (2 references per scene in two resolution).% An example snapshots of used video sequences is shown in Fig.~\ref{Snapshots}. 
 
 \begin{table}[t]
 	\centering
 	\renewcommand{\arraystretch}{1.15}
 	\caption{Test video sequences}
 	\begin{tabular}{l*{6}{c}r}
 		\hline
 		Sequence & HD & UHD & Fps & Bit Depth\\
 		\hline
 		
 		%AerialCrowd2 & \multirow{7}{*}{1920 x 1080} &  \multirow{7}{*}{3840 x 2160} & 30 &10  \\
 		{\it AerialCrowd2} & $1920\times1080$ &  $3840 \times 2160$ & 30 &10  \\
 		{\it CatRobot1 } &  $1920\times1080$ &  $3840 \times 2160$  & 60  &10\\
 		{\it CrowdRun }    &   $1920\times1080$ &  $3840 \times 2160$   & 50  &8 \\
 		{\it DaylightRoad}    &  $1920\times1080$ &  $3840 \times 2160$   & 60& 10 \\
 		{\it Drums2} &  $1920\times1080$ &  $3840 \times 2160$   & 50& 10\\
 		{\it HorseJumping} &  $1920\times1080$ &  $3840 \times 2160$   & 50&10 \\
 		{\it Sedof} &  $1920\times1080$ &  $3840 \times 2160$   & 60& 8\\
 		\hline
 	\end{tabular} \vspace{-2mm}
 	\label{table1}
 \end{table}

 In order to measure video contents diversity across the selected test sequences, \gls{si} and \gls{ti} metrics are used~\cite{itu_recommandation_1999}.
 The \gls{si} estimates the amount of spatial details whereas the \gls{ti} measures the quantity of motion in the sequence.
 In \Figure{\ref{fig:SI_TI}}, the 14 test sequences are represented under the \gls{si} \gls{ti} coordinates. The green crosses and red stars correspond to \gls{uhd} (4K) and \gls{hd} test sequences, respectively. 
 \Figure{\ref{fig:SI_TI}} shows that the down-sampled \gls{hd} sequences have equivalent \gls{ti} coordinates and significantly higher \gls{si} coordinates compared to the original \gls{uhd} sequences.
 This is due to the increased quantity of spatial details contained in the down-sampled \gls{hd} frames compared to the original \gls{uhd} (4K) frames, since the same visual information is located in a smaller frame.
 %	Indeed, the same visual information is contained in the down-sampled frames compared to the original frames, leading to an increased quantity of spatial details. 

 \begin{figure}[htbp]
 	\centerline{\includegraphics[width=1.05\linewidth]{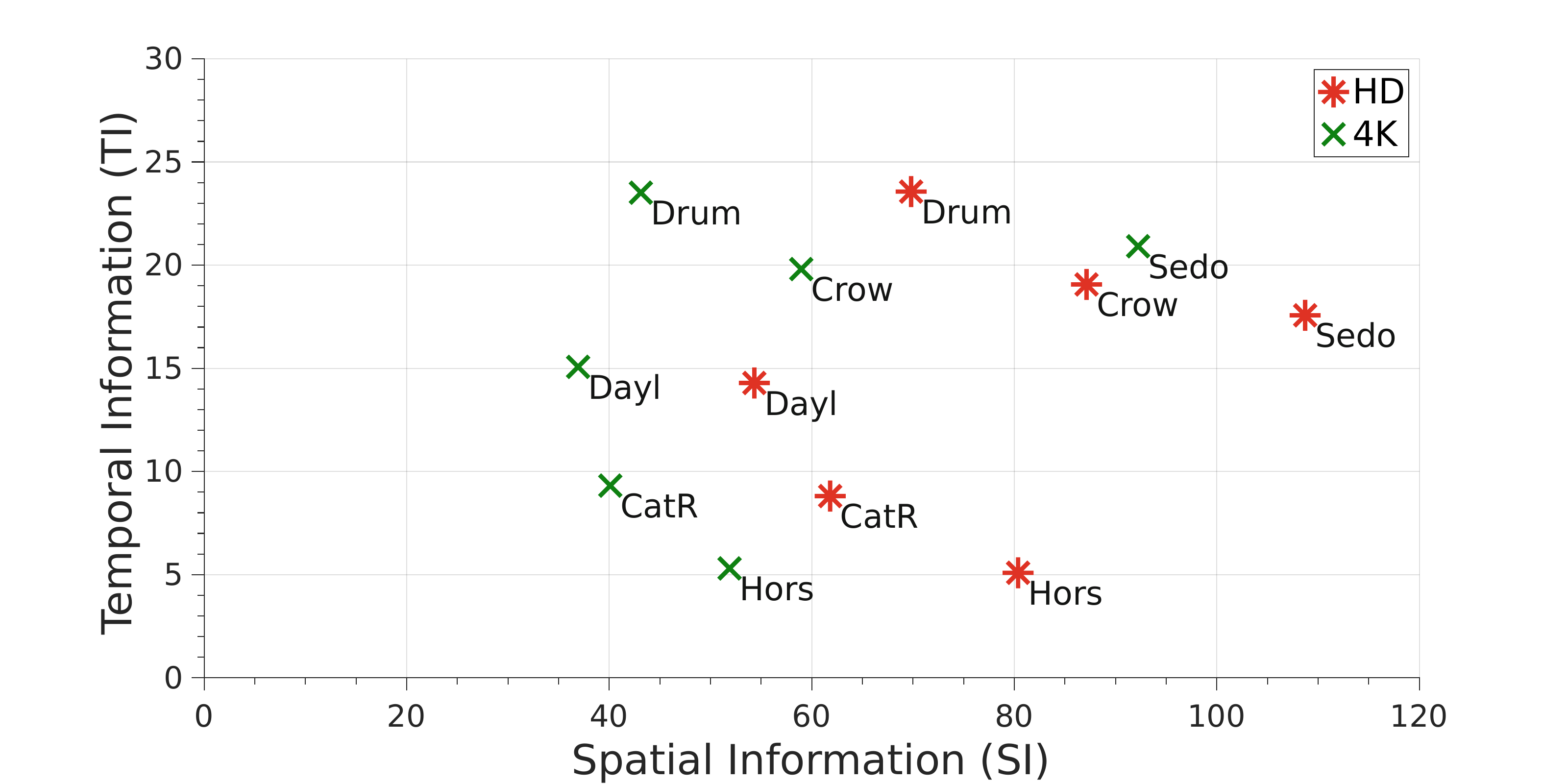}}
 	\caption{\gls{si} and \gls{ti} of the test sequences, according to the resolution.}
 	\label{fig:SI_TI}
 \end{figure}

 \subsubsection {Evaluation Procedure}
 
 In this quality assessment experiment, the Subjective Assessment Methodology for Video Quality (SAMVIQ) method was used \cite{Kozamernik}. This method has specifically been designed for multimedia content. It takes into account a range of codec types, image formats, bit-rates, temporal resolutions, etc. It has been recommended by ITU-R 6Q in 2004 \cite{Draft}. Each scene (video sequence) is presented with the following conditions: an explicit reference, a hidden reference and 10 Processed Video Sequences (PVSs). Four categories of PVS are tested (\gls{hevc}-\gls{hd}, \gls{vvc}-\gls{hd}, \gls{hevc}-\gls{uhd} and \gls{vvc}-\gls{uhd}). The button with label REF clearly identifies the explicit reference sequence. The hidden reference is identical to the explicit reference but it is not readily accessible to the subject and it is "hidden" among other stimuli. For each scene, participants were asked to evaluate the processed video sequences, given by buttons with letter labels A to K (including the hidden reference), as indicated by the protocol SAMVIQ \cite{EBU}. The conducted experiment is divided into two parts: \gls{hd} and \gls{uhd}. For an optimal visual comfort, these two parts have been done separately but using the same participants. Moreover, to prevent from visual fatigue, each part (\gls{hd} and \gls{uhd}) of the test is carried-out in two sessions.  Before each experiment, participants receive clear and deep explanations about the evaluation procedures and the used interface. Finally, all viewers scores have been collected using a dedicated Graphical User Interface (GUI), developed in compliance with the SAMVIQ recommendation.
% 
% \begin{figure}[t]
% 	\begin{minipage}[b]{1.0\linewidth}
% 		\centering
% 		\centerline{\includegraphics[width=0.86\linewidth]{fig_PCS/scanshoots_all}}
% 		%  \vspace{2.0cm}
% 		\caption{An example frames of the used video sequences (HD \& UHD).}
% 		\label{Snapshots}
% 	\end{minipage}
% \end{figure}
% 

 \section {Quality Evaluation Results} 
 \label{Experimental Results} 
 
 \subsection {Objective Evaluation Results}
 
The analysis given here is based on the four objective metrics described above: \gls{psnr}, \gls{ssim}, \gls{vmaf} and Bj{\o}ntegaard Measure (BD-BR). The objective quality results, for the whole test sequences, are shown in Fig. \ref{Objective} in the form of PSNR (top), \gls{ssim} (middle) and \gls{vmaf} (bottom) versus bit-rate plots. The solid and dotted lines represent \gls{hevc} and \gls{vvc} codec curves, respectively. It is worth noting here that the objective (i.e., \gls{psnr}, \gls{ssim}, \gls{vmaf}) and subjective (i.e., MOS) results are independently analyzed, and thus no direct correlation between them is demonstrated, since we do not develop here a new objective quality metric. Also, the \gls{psnr} results presented here are weighted through the three components ($y$, $u$, $v$}), as shown in the Equation~\eqref{psnr}. Figs. \ref{Objective} (top) illustrates the average performance in terms of \gls{psnr} metric of the whole used dataset.  According to these figures, \gls{psnr} values increase significantly when using \gls{vtm} coding tools, and consequently \gls{vtm} enables higher video quality than the \gls{hm} codec. In addition, it can be observed that \gls{vvc} codec achieves the same objective quality as \gls{hevc} while typically requiring substantially lower bit-rates. In these figures, \gls{vvc} codecs enable a bit-rate saving up to $40\%$ for some sequences as \textit{CatRobot} and \textit{DaylightRoad} in the two used formats \gls{hd} (left) and \gls{uhd} (right). 
\par The obtained results using \gls{ssim} metric, Figs. \ref{Objective} (middle), have shown also that  \gls{vvc} codecs enables a bit-rate saving up to 50\% for some test sequences, in \gls{hd} (left) and \gls{uhd} (right) formats, compared to the \gls{hevc} standard. In these figures, we can noticed that except for the \textit{CrowdRun} sequence in low bit-rate, all the used test sequences have a good quality measures: higher than $0.96$ in \gls{hd} and $0.97$ in \gls{uhd} (4K) formats. For some sequences, as \textit{AerialCrowd2} in \gls{hd} and \textit{DaylightRoad} in \gls{uhd} formats, a considerable bit-rate gains is obtained by \gls{vvc} codec while keeping the same quality measures. In other words, for the same quality at high bit-rate ($>0.99$), a bit-rate gain from $6Mbps$ to $8Mbps$ is obtained. This gain is ranking from $13Mbps$ to $21Mbps$ in \gls{uhd} (4K) format.  
\par Finally, Figs. \ref{Objective} (bottom) present the objective measurements using \gls{vmaf} metric, in \gls{hd} (left) and \gls{uhd} (right) contents. Similar to the other metrics, these figures indicate that \gls{vvc} codecs enable a bit-rate saving up to $49\%$ compared to \gls{hevc} standard. It is noticed here that we have used these metrics, while presenting the similar behaviours, in order to cover a large number of quality measurement categories, for a possible future comparisons as well as a multiple ground truth datasets.  
\par Table \ref{tablebdratepsnr} summarizes the  Bj{\o}ntegaard  measurement (BD-BR) for \gls{hd} and \gls{uhd} contents, using \gls{psnr} metric. On average, the \gls{vtm} codec enables, in average, a bit-rate savings of about 31\% and 34\% for \gls{hd} and \gls{uhd} (4K) video sequences, respectively. Moreover, using \gls{vmaf} and \gls{ssim}, the same behaviors are noticed and a significant bit-rate gains is obtained. These gains are summarized in Table~\ref{tableall}.
 
 \begin{table}[h]
 	\centering
 	\renewcommand{\arraystretch}{1.15}
 	\caption{BD-BR (\gls{psnr}) of \gls{vtm}-5 compared to the anchor HM video codec.}
 	\begin{tabular}{l*{6}{c}r}
 		\hline
 		 Sequence & BD-BR (HD) & BD-BR (UHD)\\
 		\hline
 		{\it AerialCrowd2} &    -24,5\% &-29,9\% \\
 		{\it CatRobot1}  &       -39,8\% & -41,8\% \\
 		{\it CrowdRun }    &    -27,2\% & -30,3\%  \\
 		{\it DaylightRoad}    & -38,9\% &  -42,4\% \\
 		{\it Drums2} &           -29,5\%  & -32\%\\
 		{\it HorseJumping} & -31,1\% &-32,5\%  \\
 		{\it Sedof} &              -27.7\% &  -31,9\% \\
 		\hline
 		\hline
 		\textbf{Average} & \textbf{-31.24\% }&\textbf{-34.42\%}\\
 		\hline
 	\end{tabular}
 	\label{tablebdratepsnr} \vspace{-2mm}
 \end{table}

 \subsection {Subjective Assessment Results}
 
 The first step in the results analysis is to calculate the Mean Opinion Score (MOS) for each video used in the experiment. This average is given by the equation~\ref{equation1}.

\begin{equation}
MOS_{jk} = \frac{1}{N} \sum \limits_{{i=1}}^N s_{ijk},
\label{equation1}
\end{equation}

where  $s_{ijk}$  is the score of participant $ i $ for the test video $ j $ of the sequence $ k $ and $ N $ is the number of observers.
In order to better evaluate the reliability of the obtained results, it is advisable to associate for each MOS score a confidence interval, usually at 95\%. This is given by the equation~\ref{equation2}. Scores respecting the experiment conditions must be contained in the interval $[MOS_{jk} -  IC_{jk},  MOS_{jk} + IC_{jk}]$.

\begin{equation}
IC_{jk} = 1.95 \frac{\delta_{jk}}{\sqrt{N}}, \quad  \delta_{jk} = \sqrt{\sum \limits_{{i=1}}^N \frac{(s_{ijk} - MOS_{jk})}{N}}.
\label{equation2}
\end{equation}

Before data analysis, we conducted a verification of the distribution of individual participant scores. In fact, some data may parasite the results. Thus, a filtering procedure was applied to obtained results based on the annex 2 of the ITU-R BT.500 recommendation~\cite{Draft}.  In our outlier detection verification, a correlation index greater (Minimum between Pearson and Spearman correlations) than or equal to 0.75 is considered as valid for the acceptance of the viewer's scores; otherwise, the viewer is considered as an outlier. Following this screening process, 8 subjects (6 in \gls{hd} and 2 in \gls{uhd} test) were discarded. Consequently, only 36 viewers scores are retained. 

Table~\ref{tablebdrate} summarises the BD-Rate gains obtained by \gls{vvc} relative to \gls{hevc}, considering \gls{MOS}. As shown in this table, \gls{vtm} outperforms the \gls{hevc} for the whole used sequences. A bit-rate savings of about 37\% and 40\% are obtained for \gls{hd} and \gls{uhd} (4K) video sequences, respectively

\begin{figure*}[!tbp]
  \centering
\begin{minipage}[b]{0.3\textwidth}
    \includegraphics[width=\textwidth]{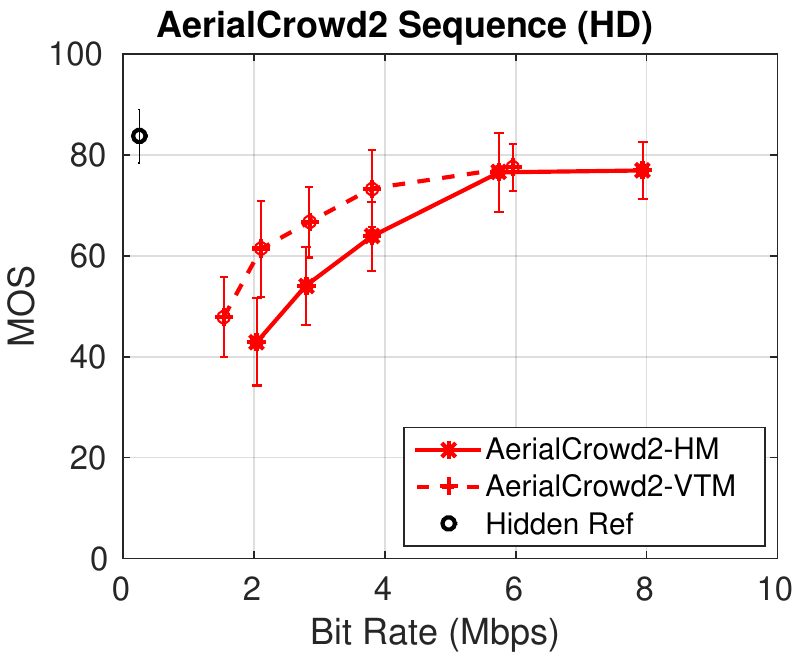}
  \end{minipage}
  \begin{minipage}[b]{0.3\textwidth}
    \includegraphics[width=\textwidth]{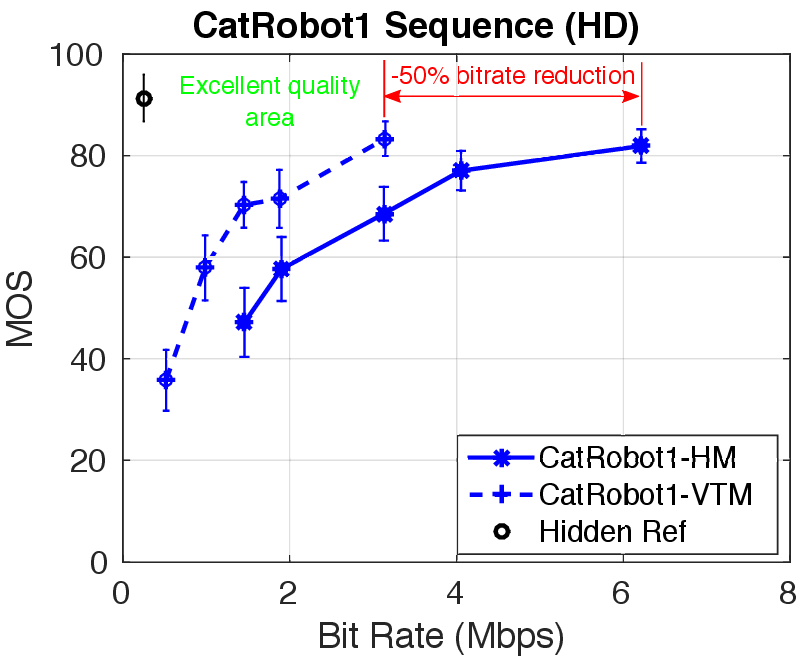}
  \end{minipage}
  \begin{minipage}[b]{0.3\textwidth}
    \includegraphics[width=\textwidth]{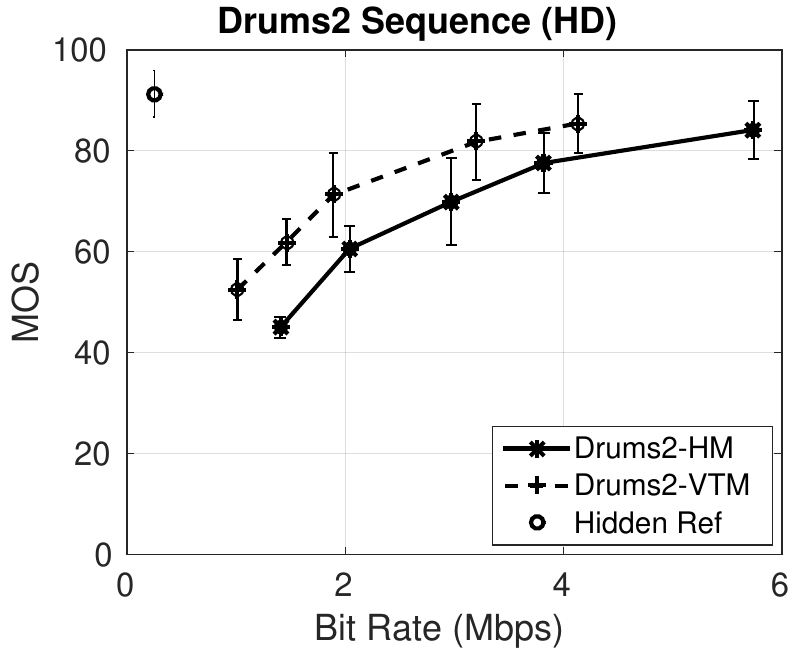}
  \end{minipage}
  \begin{minipage}[b]{0.3\textwidth}
    \includegraphics[width=\textwidth]{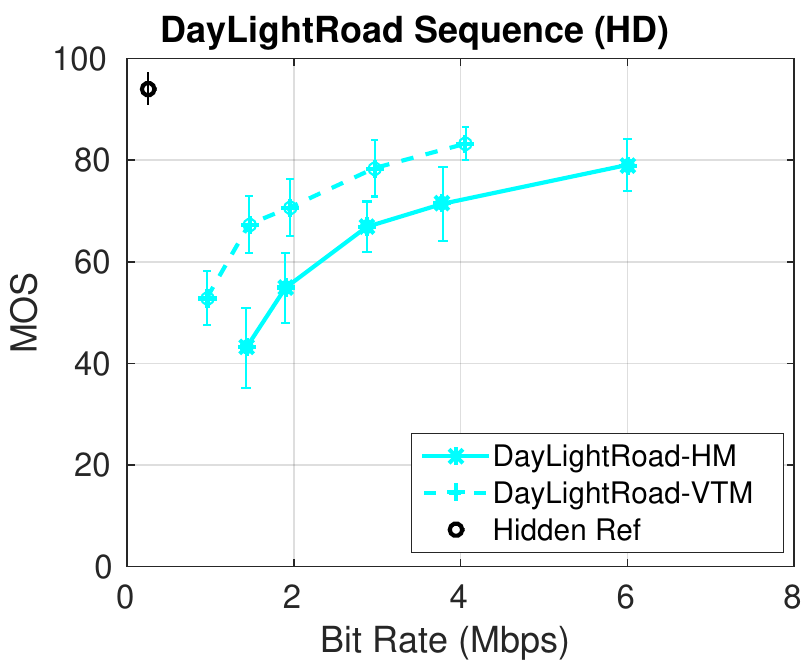}
  \end{minipage}
  \begin{minipage}[b]{0.3\textwidth}
    \includegraphics[width=\textwidth]{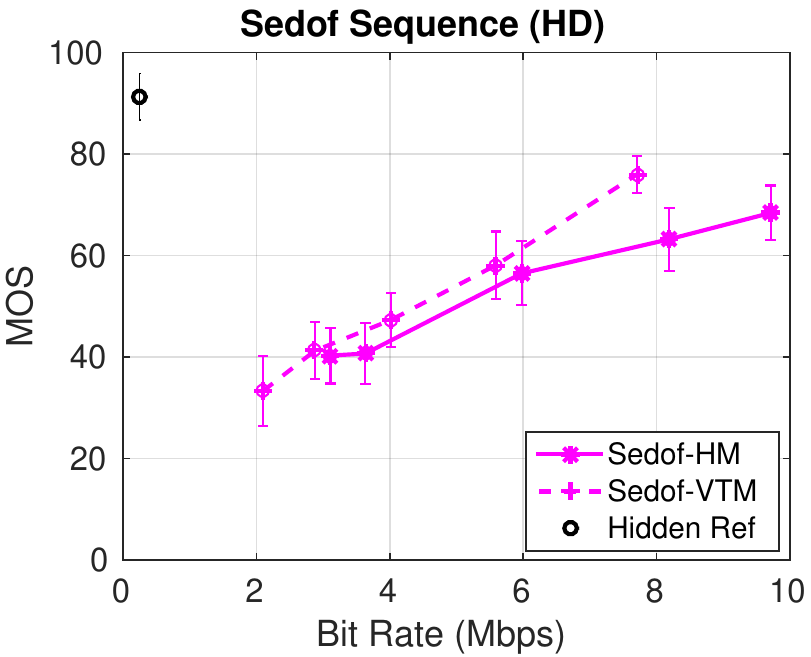}
  \end{minipage}
  \begin{minipage}[b]{0.3\textwidth}
    \includegraphics[width=\textwidth]{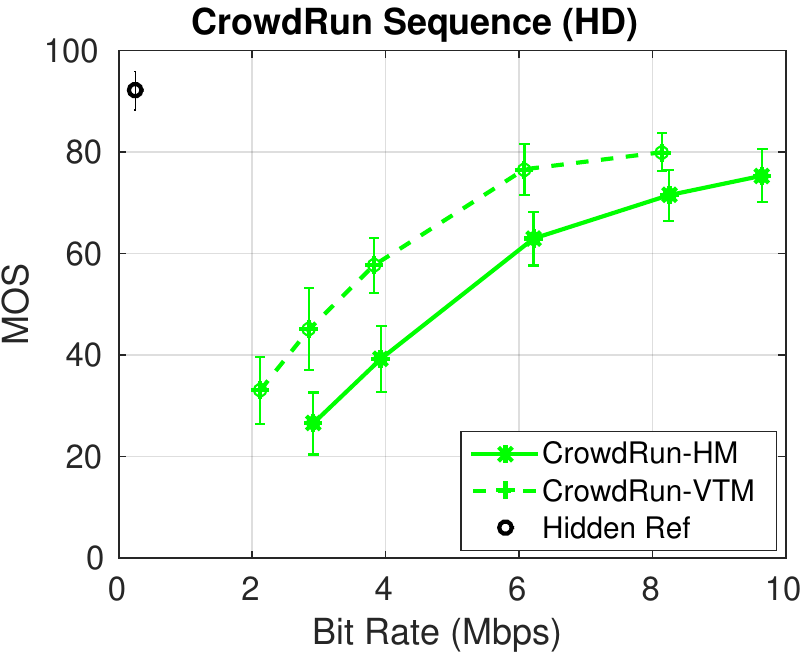}
  \end{minipage}
    \begin{minipage}[b]{0.3\textwidth}
    \includegraphics[width=\textwidth]{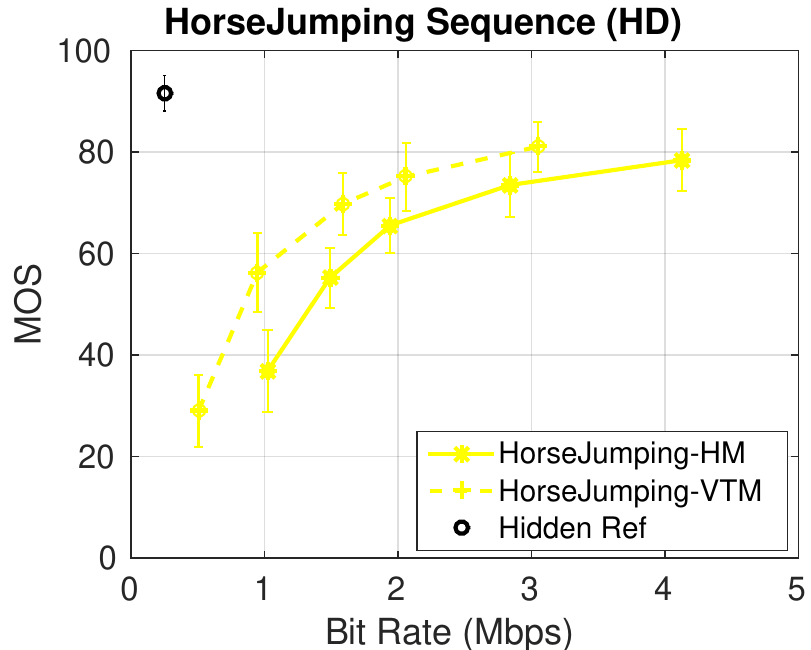}
  \end{minipage}
   \caption{MOS-based comparison, with associated 95\% confidence intervals, in \gls{hd} format.}
   \label{MOS1}
\end{figure*}

\begin{figure*}[!tbp]
  \centering
\begin{minipage}[b]{0.3\textwidth}
    \includegraphics[width=\textwidth]{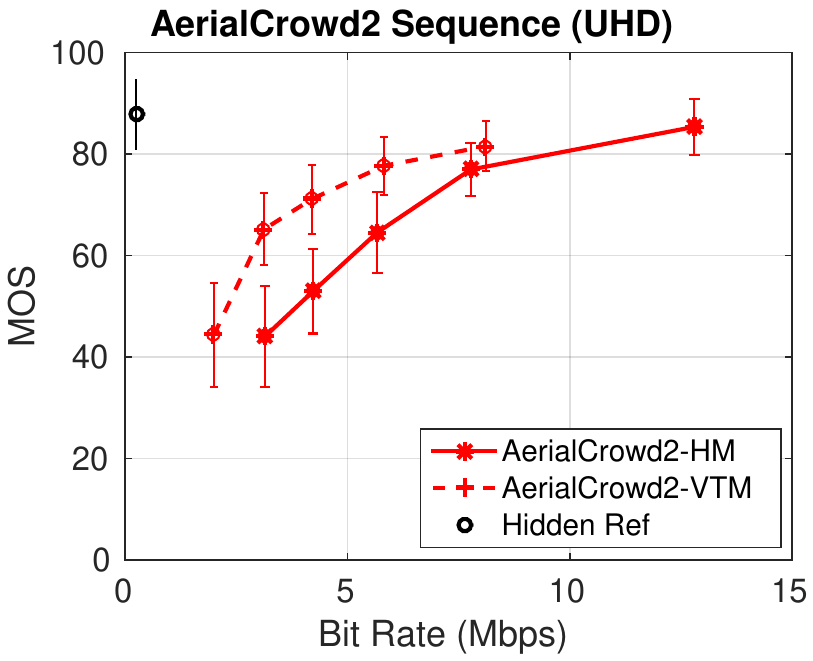}
  \end{minipage}
  \begin{minipage}[b]{0.3\textwidth}
    \includegraphics[width=\textwidth]{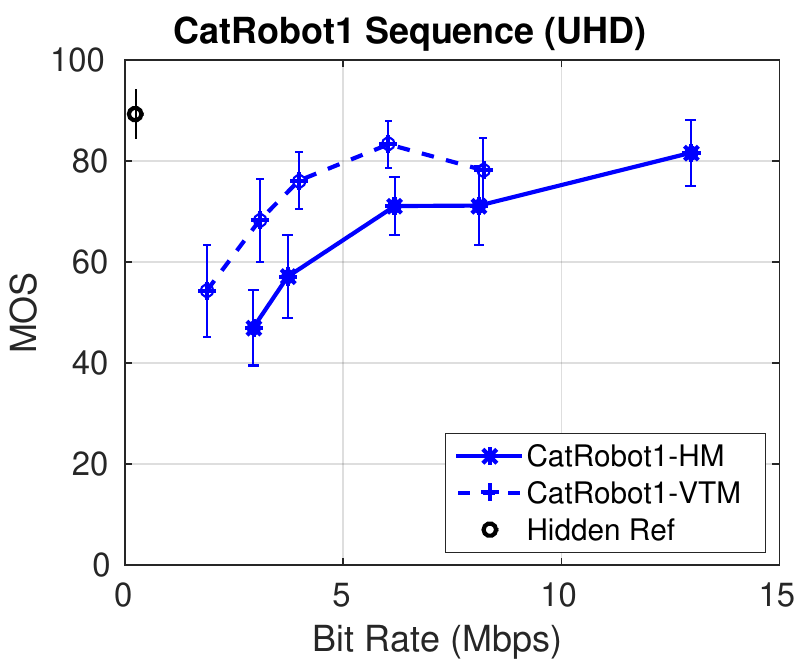}
  \end{minipage}
  \begin{minipage}[b]{0.3\textwidth}
    \includegraphics[width=\textwidth]{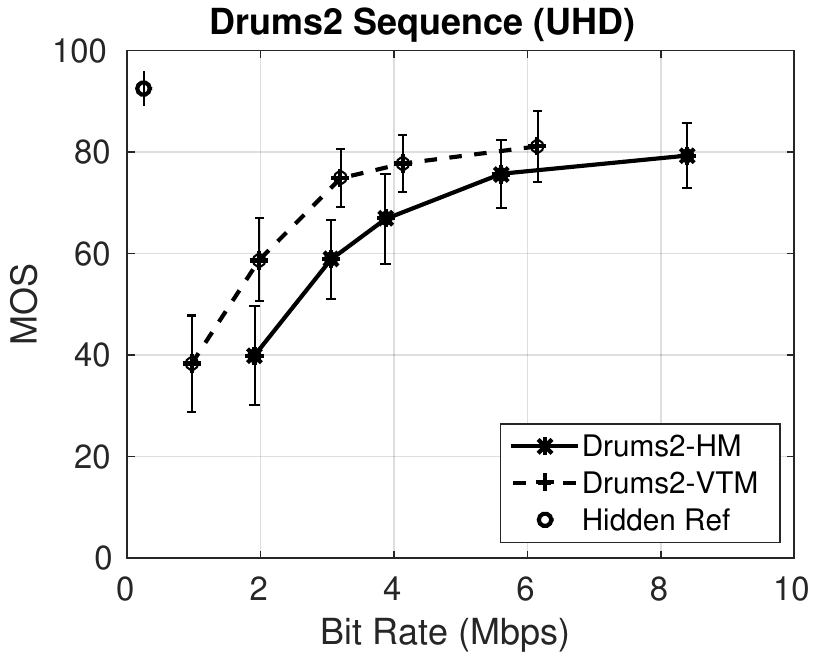}
  \end{minipage}
  \begin{minipage}[b]{0.3\textwidth}
    \includegraphics[width=\textwidth]{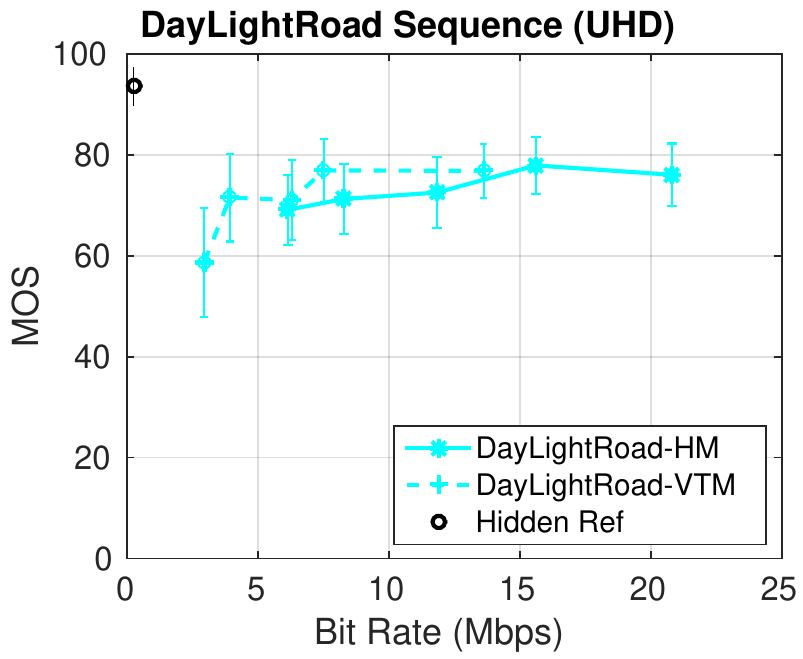}
  \end{minipage}
  \begin{minipage}[b]{0.3\textwidth}
    \includegraphics[width=\textwidth]{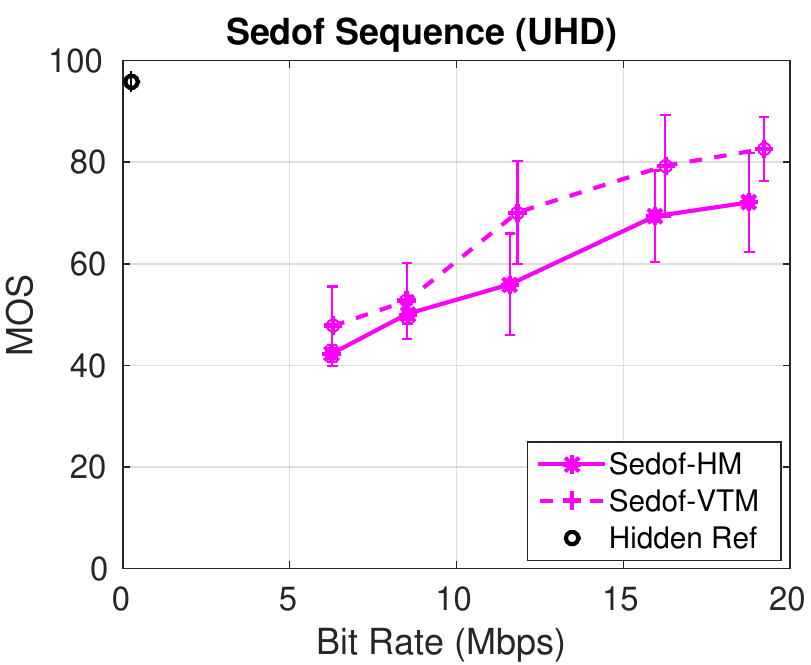}
  \end{minipage}
  \begin{minipage}[b]{0.3\textwidth}
    \includegraphics[width=\textwidth]{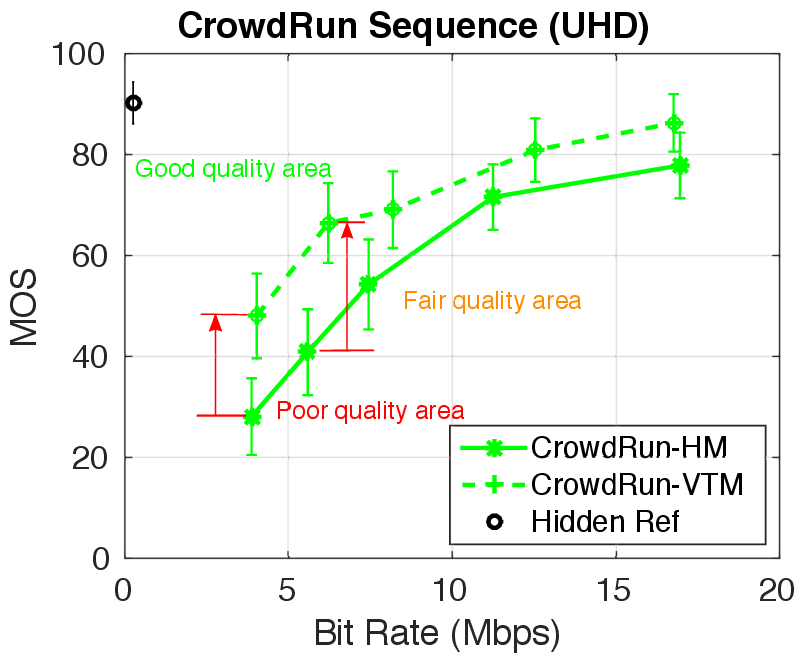}
  \end{minipage}
    \begin{minipage}[b]{0.3\textwidth}
    \includegraphics[width=\textwidth]{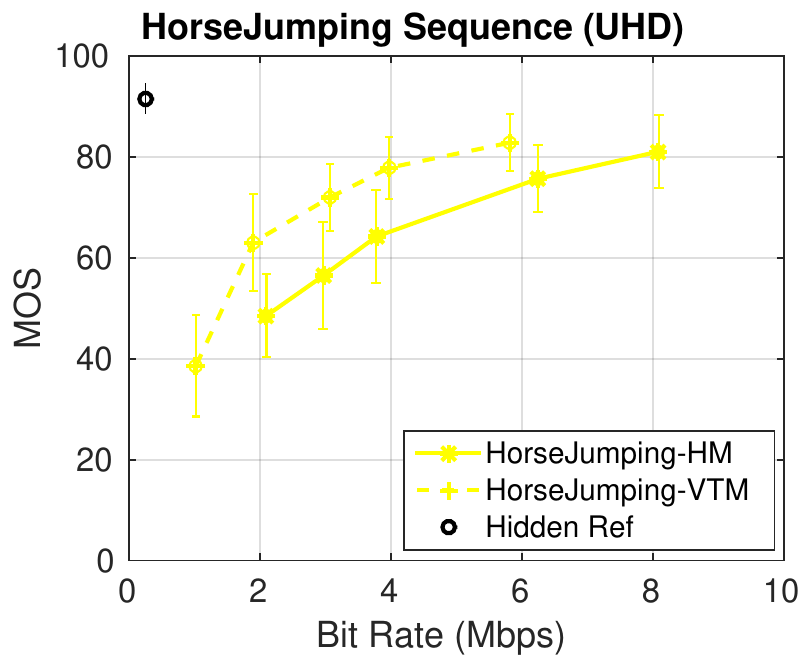}
  \end{minipage}
   \caption{MOS-based comparison, with associated 95\% confidence intervals, \gls{uhd} (4K) format.}
   \label{MOS2}
\end{figure*}

The subjective evaluation results, for the whole test sequences are shown in Figs. \ref{MOS1} \& \ref{MOS2} in the form of MOS versus bit-rate plots, for \gls{hd} and \gls{uhd} (4K) formats, respectively. The solid and dotted lines represent  \gls{hevc} and \gls{vvc} codec curves, respectively. The associated confidence intervals are displayed for each MOS test point. In these figures, \gls{vtm} codec enables a higher \gls{MOS} score and consequently a higher video quality, compared to \gls{hevc} reference software. As we can see, a considerable gains in terms of bit-rate savings can be noticed with the same perceived quality. For \gls{hd} format , a bit-rate saving of -50\% and -40\% is obtained in the ``Excellent'' and ``Good'' quality area, respectively. For \gls{uhd} format, a bite-rate savings of about -50\% is obtained, with the same perceived quality, for the sequence {\it CatRobot1}. Reciprocally, for the same bit-rate, a considerable quality enhancement is obtained by \gls{vvc} compared to \gls{hevc} standard. In fact, at low bit-rate, the sequence ``CrowdRun'' is judged as ``Fair'' quality using \gls{vvc} while it is judged as ``Poor'' quality using \gls{hevc} standard. In addition, this quality is enhanced from ``Fair'' to ``Good'' quality area using \gls{vvc}. The same behavior can be noticed for the other video sequences, depending on the considered bit-rates and the used video format.

\begin{figure*}[!t]
  \centering
\begin{minipage}[b]{0.32\textwidth}
    \includegraphics[width=\textwidth]{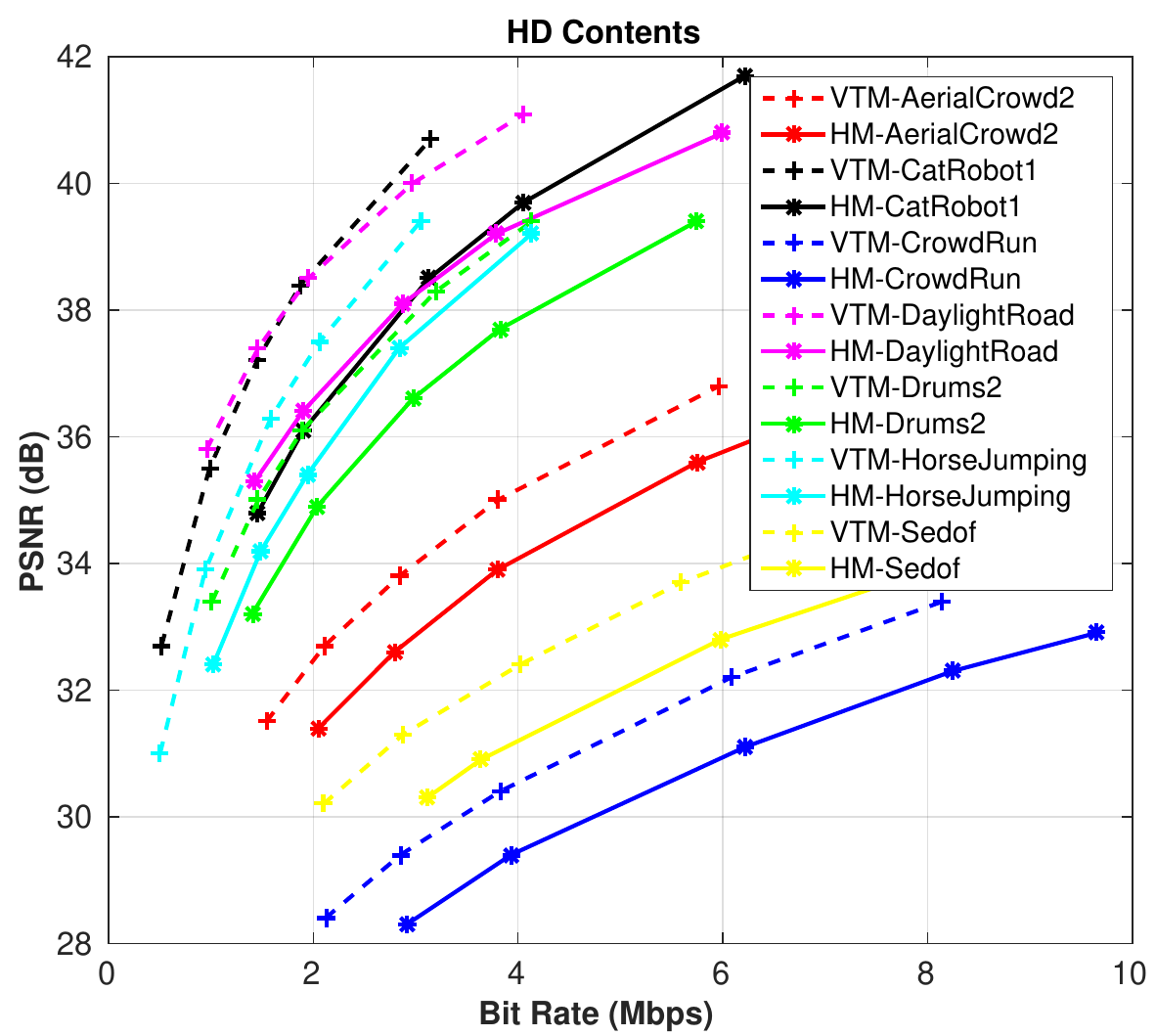}
  \end{minipage}
\vspace{0.6cm}
  \begin{minipage}[b]{0.32\textwidth}
    \includegraphics[width=\textwidth]{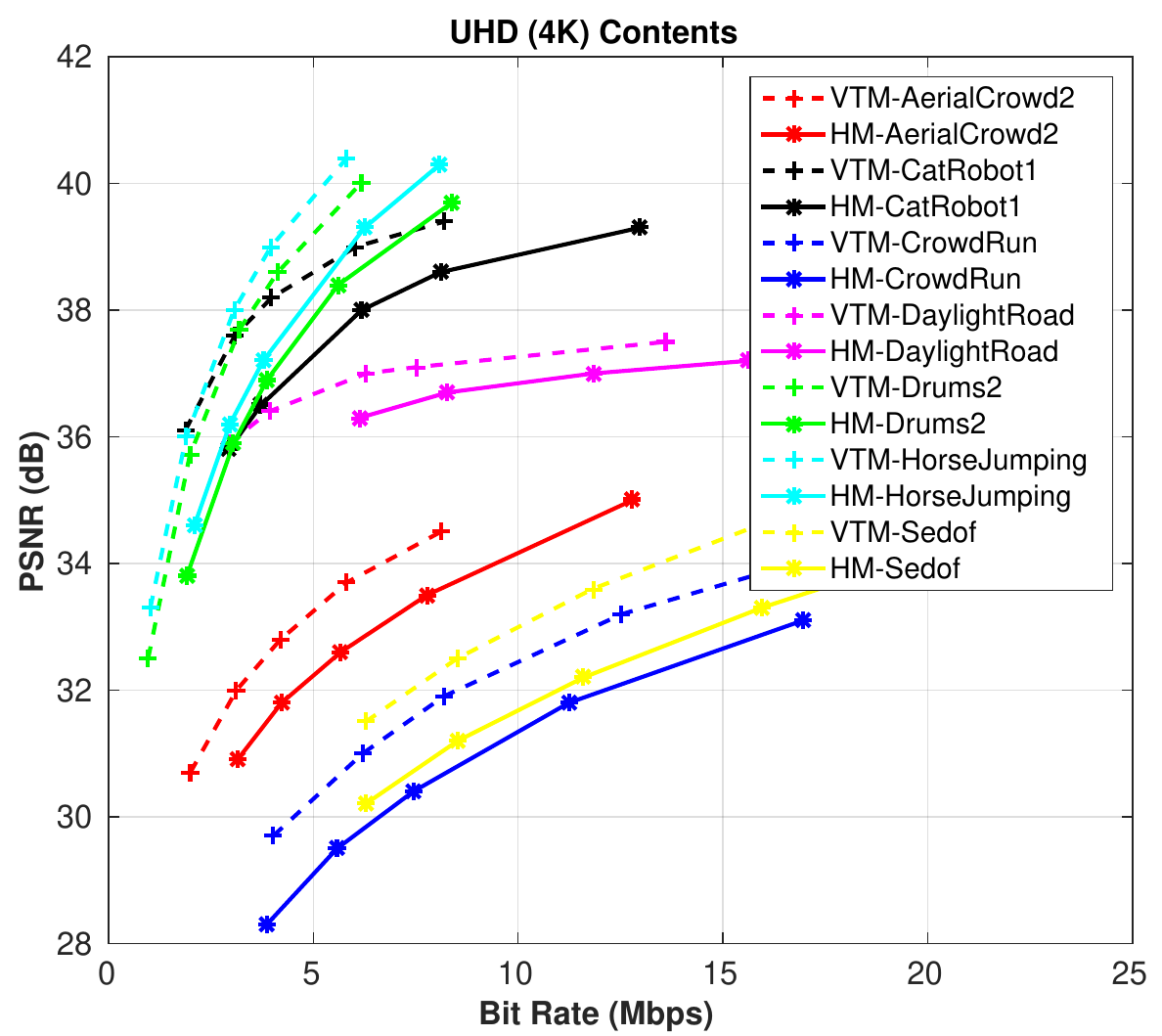}
  \end{minipage}
%\vspace{0.5cm}
\begin{minipage}[b]{0.32\textwidth}
    \includegraphics[width=\textwidth]{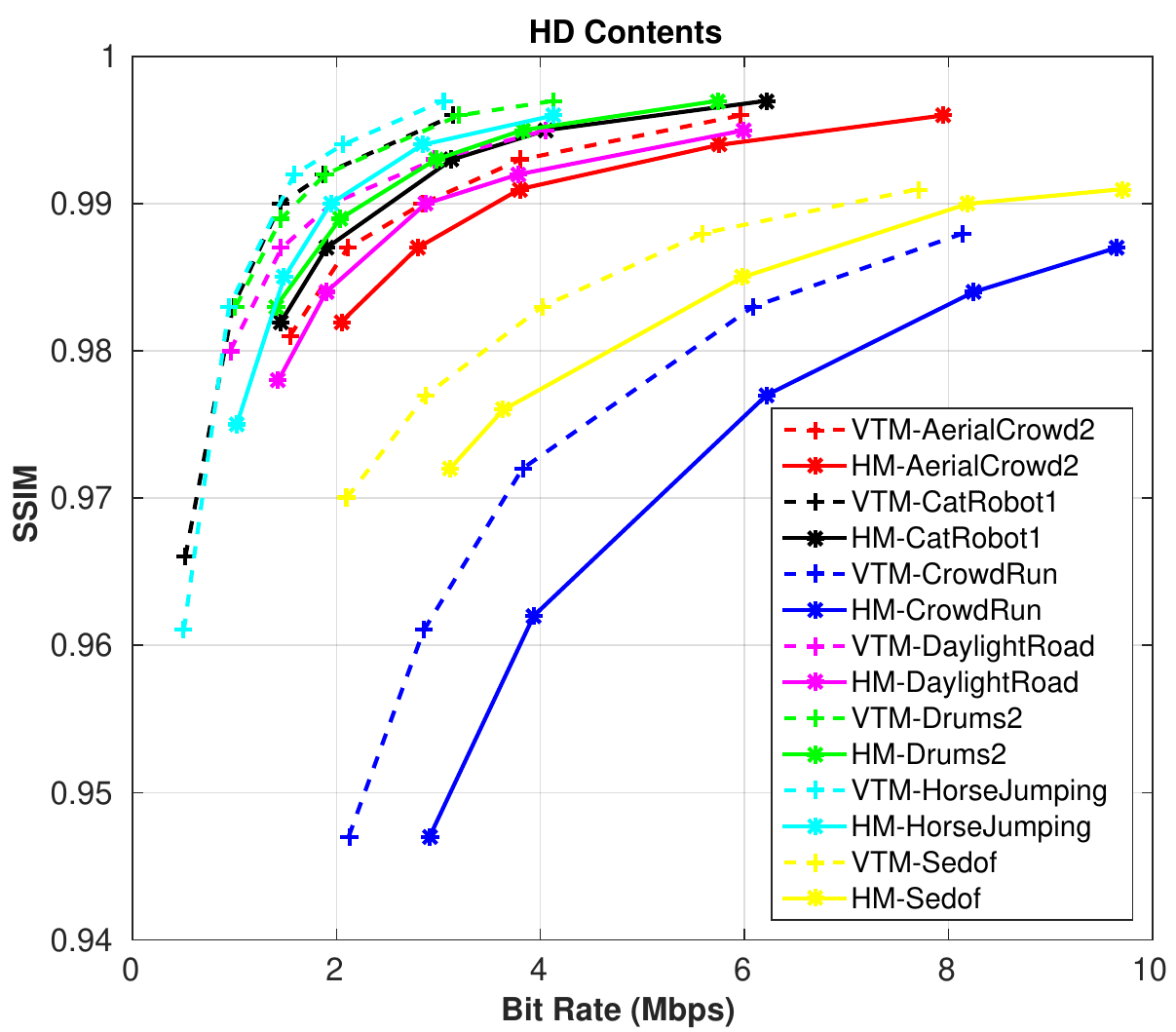}
  \end{minipage}
  \begin{minipage}[b]{0.32\textwidth}
    \includegraphics[width=\textwidth]{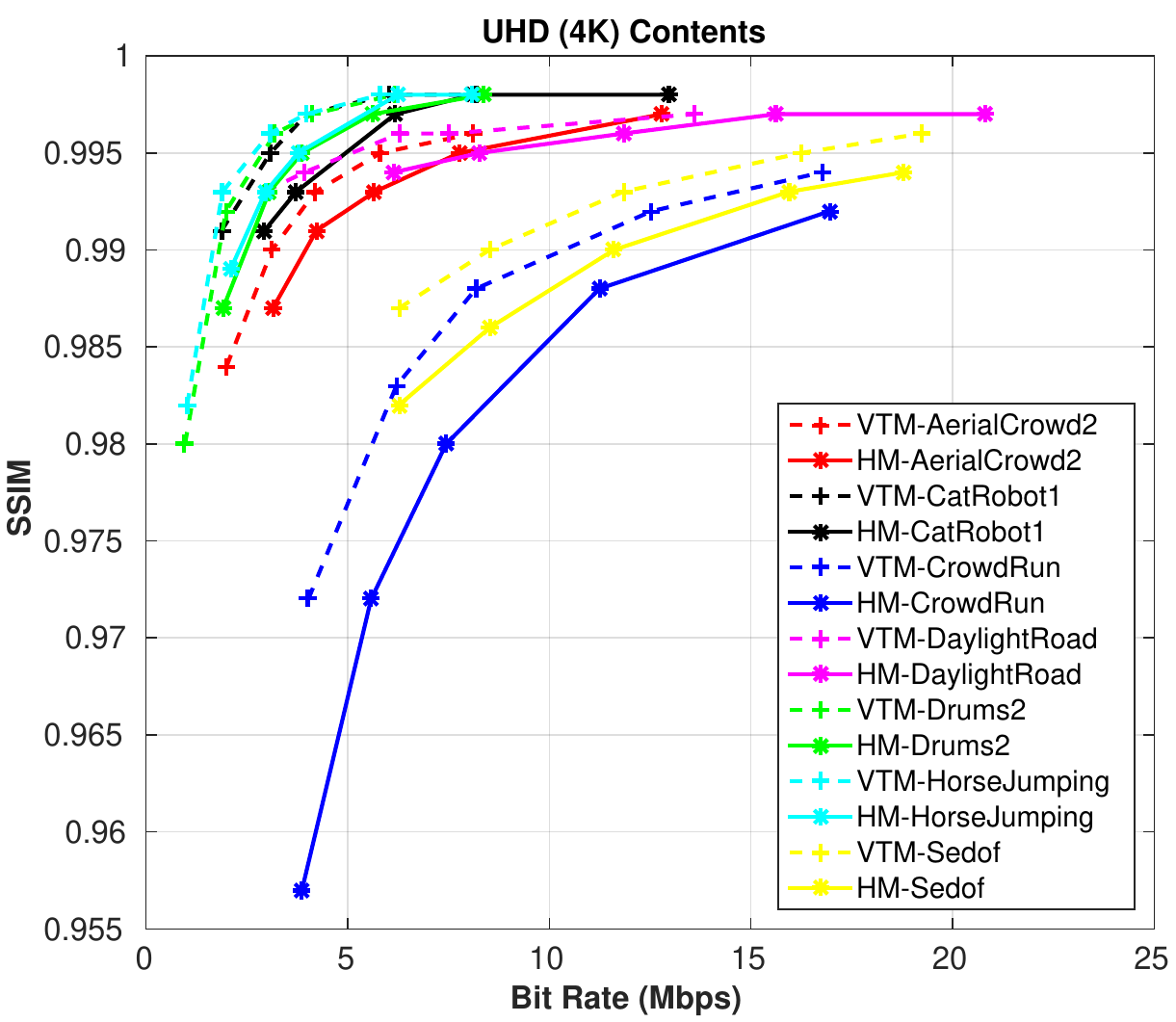}
  \end{minipage}
\begin{minipage}[b]{0.32\textwidth}
    \includegraphics[width=\textwidth]{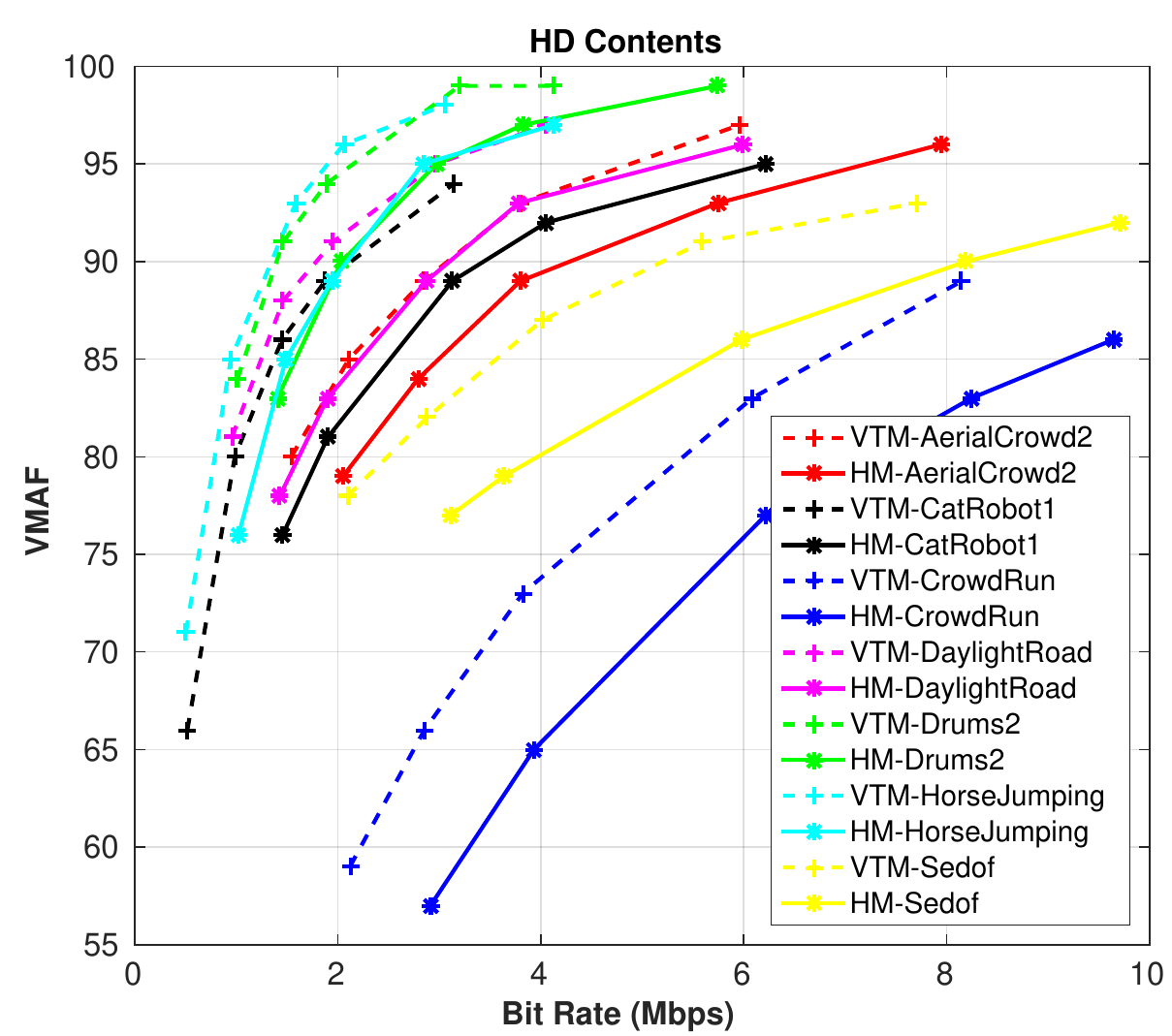}
  \end{minipage}
  \begin{minipage}[b]{0.32\textwidth}
    \includegraphics[width=\textwidth]{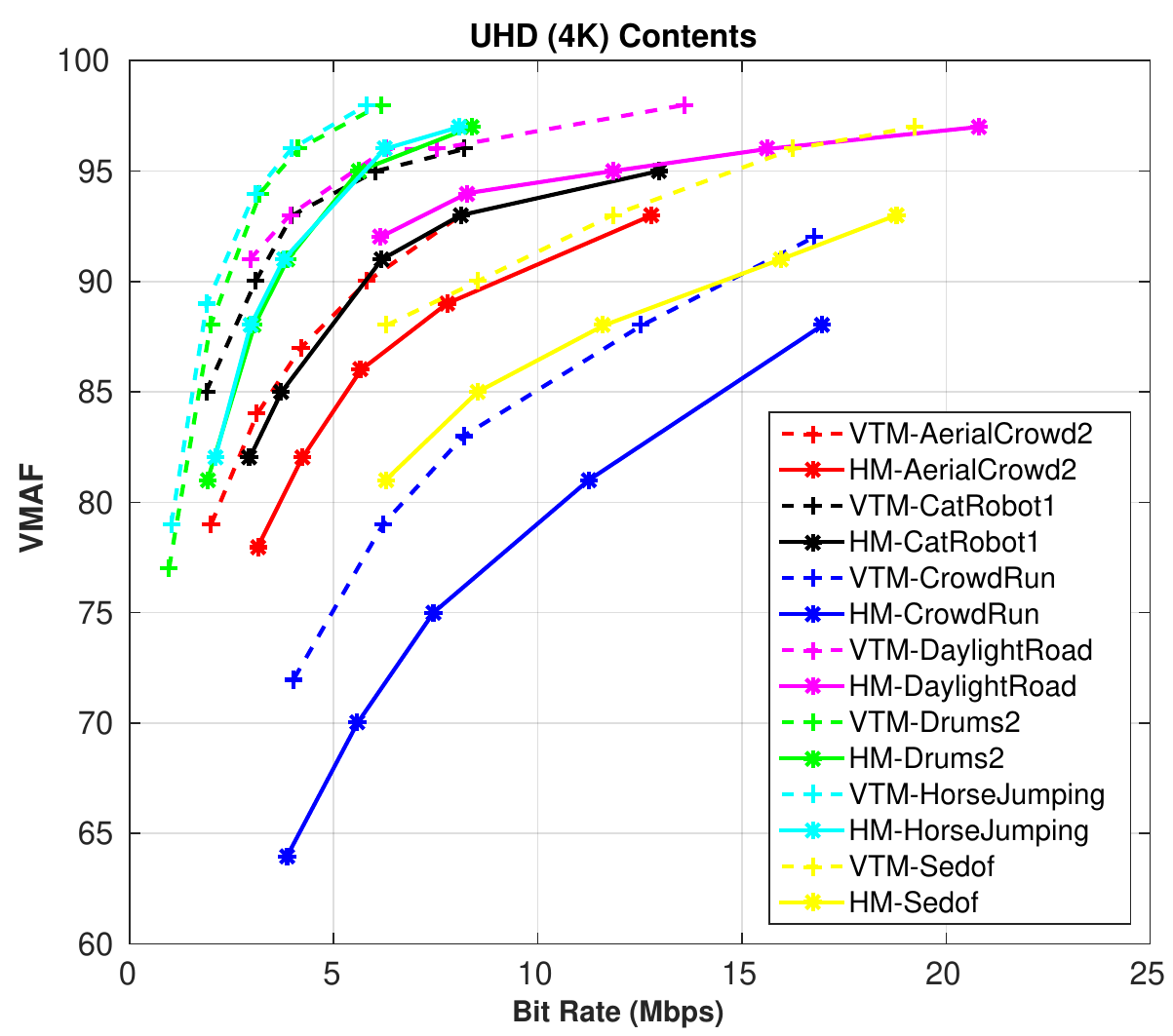}
  \end{minipage}
   \caption{Objective-based comparison, using three metrics: \gls{psnr} (top), \gls{ssim} (middle) and VMAF (bottom), for the whole used datasets in \gls{hd} and \gls{uhd} (2160p) formats.}
   \label{Objective}
\end{figure*}

 \begin{table}[h]
 	\centering
 	\renewcommand{\arraystretch}{1.15}
 	\caption{BD-BR (MOS) of \gls{vtm} compared to the anchor \gls{hm}-16.20 video codec.}
 	\begin{tabular}{l*{6}{c}r}
 		\hline
 		Sequence & BD-BR (HD) & BD-BR (UHD)\\
 		\hline
 		
 		{\it AerialCrowd2} & -33\% &-41\% \\
 		{\it CatRobot1}  & -48\% & -47\% \\
 		{\it CrowdRun }    & -35\% & -35\%  \\
 		{\it DaylightRoad}    & -49\% &  -48\% \\
 		{\it Drums2} & -35\% & -39\% \\
 		{\it HorseJumping} & -38\% &-45\%  \\
 		{\it Sedof} & -19\% &  -21\% \\
 		\hline
 		\hline
 		\textbf{Average} & \textbf{-37\% }&\textbf{-40\%}\\
 		\hline
 	\end{tabular}
 	\label{tablebdrate} \vspace{-2mm}
 \end{table}

 As a conclusion, for \gls{hd} resolution, a bit-rate saving of 31\% and 35\% can been achieved with the \gls{vtm} in terms of \gls{psnr} and \gls{vmaf} metrics, respectively. This gain exceeds 40\% for the \gls{uhd} resolution, using \gls{vmaf} metric. For the subjective comparison, the obtained gains is ranging between 37\% and 40\% for \gls{hd} and \gls{uhd} resolutions, respectively, as summarized in Table~\ref{tableall}. For the same bit-rate range, the highest bit-rate saving for \gls{hd} contents is obtained by MOS-based BD-rate (-37\%) while this gains for \gls{uhd} contents is obtained by \gls{vmaf}-based BD-rate (-40.44\%). 

\subsection {Statistical Analysis}

A statistical analysis was performed using the Analyse of Variance (ANOVA) approach \cite{anova}. Bit-rate, Content, Resolution and Codec are used as independent variables and MOS is used as dependent variable. The null hypothesis in this case would be that the\gls{vvc} test points have the same quality as the \gls{hevc} test point while the alternate hypothesis is that the \gls{vvc} test points do not have the same perceived quality as the \gls{hevc} test point \cite{anova}. Results have shown that only codec parameter (\gls{vtm}, \gls{hm}) has a significant influence on the subjects scores, with \textit{p-value $< 0.0001$}~\footnote{a factor is considered influencing if \textit{p-value $< 0.05$}}. Subjectively speaking, \gls{vvc} standard enables a significant visual quality improvement regardless the used bit-rate, resolution and video content.

 \begin{table}[h]
\centering
\renewcommand{\arraystretch}{1}
\caption{Bit-rate Savings of VTM over HEVC standard.}
\begin{tabular}{l*{6}{c}r}

{Resolution} &{ \gls{psnr}} & { VMAF} & { SSIM} & {MOS}\\
\hline

{HD} & -31.24 \% & -35.18 \%& -33\% & \textbf{-37\%}  \\
\hline
{UHD}  & -34.42\% & \textbf{-40.44\%} & -38\% & -40\%\\

\hline
\end{tabular}
\label{tableall} \vspace{-2mm}
\end{table}

\section{VVC Coding Time and Complexity Repartition}
%\section{Codecs Time and Complexity Repartition}
\label{complexity}

This section provides an analysis of times and complexity repartitions, for both encoding and decoding processes. In order to highlight the principal evolutions from one reference software to another, the \gls{vtm} and \gls{hm} results are presented simultaneously.

\subsection{Platform and Configuration Parameters}\label{subsec:exp_setup}
%
%\begin{table}[h]
%	\centering
%	\caption{Configuration parameters for \gls{hm}-16.20 and \gls{vtm}-5.0 main profiles, in \gls{ra} configuration. \red{Attention aux Accronymes.}}
%	\label{tb:configurations}
%\begin{adjustbox}{max width=1\columnwidth}
%	\begin{tabular}{l|l|l}
%		\textbf{Parameter}            &   \textbf{\gls{hm}-16.20 main}    & \textbf{\gls{vtm}-5.0 main} \\\midrule
%		\textbf{\gls{ctu} size}    &  64     &  128  \\
%		\textbf{\gls{qt} max depth}   &  4      &  4    \\
%		\textbf{\gls{mtt} max depth}  &  0      &  3    \\
%		\textbf{Transform Types}      &  DCT-II, DST-VII &  DCT-II, DST-VII,  DCT-VIII \\
%		\textbf{Max \gls{tu} size}    &  32     & 64    \\
%		\textbf{Loop Filters}         &  DBF, SAO     & DBF, SAO, ALF \\
%		\textbf{Search Type}          &  TZ  &  TZ   \\
%		\textbf{Search Range}         &  384 &  384 \\
%		\textbf{N. Ref. Pictures}     &  5   &  5 \\
%		\textbf{Entropy Coding}       &  CABAC    &  CABAC \\
%		\textbf{Internal Bit Depth }  &  10  &  10 
%	\end{tabular}
%\end{adjustbox}
%\end{table}
%
%
%In this work, the encoding are carried out with the main profiles of the \gls{hm}-16.20 and \gls{vtm}-5.0 in \gls{ra} configuration. The profiles configuration parameters are summarized in \Table{\ref{tb:configurations}}.
%As detailed in Section~??, the parameters that differ from \gls{hm}-16.20 to \gls{vtm}-5.0 include \gls{ctu} size, allowed partition modes, Max \gls{tu} size and enabled loop filters.  

The experiments are carried-out on the same 14 sequences described in Section~\ref{subsec:test_video}.
%\blue{In order to analyze the impact of \gls{qp} on encoding and decoding times, 2 \gls{qp} values per resolution are retained: \gls{qp}=27 and \gls{qp}=32 for \gls{hd} sequences, \gls{qp}=32 and \gls{qp}=37 for \gls{uhd} sequences.
%The \gls{qp}=37 generates bit-rates considered too small to be representative of real use-cases for \gls{hd} sequences, and \gls{qp}=27 generates bit-rates considered too large for \gls{uhd} sequences.}
%\blue{For this reason, these two \gls{qp} values are discarded in the following experiments.}
In order to evaluate the impact of \gls{qp} on encoding and decoding times, 2 \gls{qp} values per resolution are selected.
Encodings of \gls{hd} sequences with \gls{qp}=37 generate bit-rates considered too small to be representative of real use-cases, and bit-rates generated by \gls{uhd} sequences encodings with \gls{qp}=27 are considered too large.
For this reason, the aforementioned \gls{qp} values are discarded in the following experiments.
Finally, four \gls{qp} values are retained in the following experiments: \gls{qp}=27 and \gls{qp}=32, for \gls{hd} sequences, and \gls{qp}=32 and \gls{qp}=37, for \gls{uhd} (4K) sequences.

\begin{table}[th]
	\centering
	\caption{Platform setup for complexity analysis}
	\label{tb:platform}
\begin{adjustbox}{max width=1\columnwidth}
	\begin{tabular}{l|l}
		\textbf{\gls{cpu}}                  & Intel Xeon CPU E5-2603 v4\\
		\textbf{Clock Rate}                 & 1.70~GHz \\
		\textbf{Memory}                     & 8 Gb \\
		\textbf{Compiler}                   & gcc 5.4.0 \\
		\textbf{Operating System}           & Linux 4.4.0-127-generic
	\end{tabular}
\end{adjustbox}
\end{table}

\Table{\ref{tb:platform}} details the platform setup for complexity analysis. The experiments on codecs time and complexity repartition are carried-out on an Intel(R) Xeon(R) CPU E5-2603 v4, a \gls{cpu} clocked at 1.70~GHz.
The \gls{hm} and \gls{vtm} are both built with gcc compiler version~5.4.0, under Linux version 4.4.0-127-generic as distributed in Ubuntu-16.04.
%Contrary to execution time, that depends among others on memory accesses or \gls{cpu} frequency, the insight of the encoding complexity repartition given by Callgrind is nearly constant regardless of the execution platform.

By default, the \gls{vtm} features low-level optimizations, such as \gls{simd} optimizations, at both encoder and decoder sides. \gls{simd} describes computers that perform the same operation on multiple data simultaneously, with a single instruction.
%By default, \gls{simd} optimizations are enabled at both encoder and decoder sides.
To assess the time consumption of the \gls{vtm} without low-level optimizations, the \gls{simd} optimizations at both encoder and decoder sides are disabled in the following. This also enables a fair comparison with  the \gls{hm} software that does not include any l\gls{simd} optimisation.  The speed-up offered by \gls{simd} optimizations is nonetheless measured and further discussed.

The repartition of encoding and decoding complexities, presented in upcoming Sections~\ref{subsec:encoding_complexity} and \ref{subsec:decoding_complexity}, are obtained by running the executables with Callgrind~\cite{noauthor_callgrind_nodate}.
Callgrind is the Valgrind profiling tool that records the call history of program functions as a call graph.
By default, the data collected includes the number of instructions executed, the calling/callee relation between functions and the number of calls.
Contrary to execution time, that depends among others on memory accesses or \gls{cpu} frequency, the insight of the complexity repartition given by Callgrind is nearly constant regardless of the execution platform.

\subsection{Encoders Analysis}\label{subsec:encoding_times_complexity}

\subsubsection{Encoding Time}\leavevmode\par \label{subsec:encoding_times}

\begin{table}[t]
	\centering
	\renewcommand{\arraystretch}{1.15}
	\caption{Factor between encoding time and real time (x1000), for both \gls{vtm}-5.0 and \gls{hm}-16.20 in \gls{ra} configuration, according to the test sequence and \gls{qp} value.}
	\label{tb:temps_encodage}
	\begin{adjustbox}{max width=1\columnwidth}
		%		\begin{adjustbox}{max width=1\textwidth}
		\begin{tabular}{@{}l||cc|cc||cc|cc@{}}
			
			\multicolumn{1}{c}{}
			&\multicolumn{4}{c}{\textbf{\gls{hd}}}
			&\multicolumn{4}{c}{\textbf {\gls{uhd}} }  \\\midrule
			
			\textbf{Encoder} 
			&\multicolumn{2}{c|}{\textbf{\gls{hm}-16.20}}
			&\multicolumn{2}{c||}{\textbf{\gls{vtm}-5.0}}  
			&\multicolumn{2}{c|}{\textbf{\gls{hm}-16.20}}
			&\multicolumn{2}{c}{\textbf{\gls{vtm}-5.0}} \\\midrule
			
			\textbf{QP} & 
			\textbf{27}&  
			\textbf{32} &
			\textbf{27}&  
			\textbf{32} &
			\textbf{32}&  
			\textbf{37} &
			\textbf{32}&  
			\textbf{37}    \\\midrule%\midrule
			
\emph{AerialCrowd2}& 1.1  & 0.9  & 22  & 15    & 3.9  & 3.4    & 64   & 41    \\
\emph{CatRobot1} &   2.0  & 1.8  & 33  & 26    & 7.2  & 6.7   & 102  & 68    \\
\emph{CrowdRun} &    2.4  & 2.1  & 49  & 39    & 8.6  & 7.6  & 179  & 124     \\
\emph{DaylightRoad}& 2.3  & 2.0  & 44  & 32    & 8.1  & 7.3   & 132  & 84     \\
\emph{Drums2} &      2.1  & 1.9  & 42  & 32    & 7.4  & 6.7   & 107  & 72       \\
\emph{HorseJumping}& 1.4  & 1.3  & 23  & 17    & 5.4  & 5.0   & 57   & 38    \\
\emph{Sedof}&        2.2  & 1.9  & 40  & 28    & 7.9  & 7.1  & 129  & 85     \\ \midrule %\midrule
\textbf{\emph{Mean }} &       \textbf{1.9}  & \textbf{1.7}  & \textbf{36.2}  & \textbf{26.9}    & \textbf{6.9}  & \textbf{6.3}   & \textbf{110.3 } & \textbf{73.2}   \\ %\midrule
\textbf{\emph{Std Dev}} &     \textbf{0.4}  & \textbf{0.4}  & \textbf{9.8} & \textbf{7.9}   & \textbf{1.5}  & \textbf{1.4}   & \textbf{38.9}   & \textbf{27.3} \\ \midrule
			
%			\begin{tabular}{@{}r||c|c||c|c@{}}
%			\multicolumn{1}{c}{}&\multicolumn{2}{c}{\textbf{\gls{hd}}}
%			&\multicolumn{2}{c}{\textbf {4K} }  \\\midrule
%			\textbf{Sequence name} & 
%			\textbf{\gls{hm}-16.20}&  
%			\textbf{\gls{vtm}-5.0} &
%			\textbf{\gls{hm}-16.20}&
%			\textbf{\gls{vtm}-5.0}    \\\midrule\midrule
%			
%			\emph{AerialCrowd2}&  0.9  & 15   & 3.9     & 64    \\
%			\emph{CatRobot1} &    1.8  & 26   & 7.2    & 102    \\
%			\emph{CrowdRun} &     2.1  & 39   & 8.6   & 179     \\
%			\emph{DaylightRoad}&  2.0  & 32   & 8.1    & 132     \\
%			\emph{Drums2} &       1.9  & 32   & 7.4    & 107       \\
%			\emph{HorseJumping}&  1.3  & 17   & 5.4    & 57    \\
%			\emph{Sedof}&         1.9  & 28   & 7.9   & 129     \\ \midrule\midrule
%			\emph{Mean } &        1.8  & 27   & 6.9    & 110   \\ \midrule
%			\emph{Std Dev} &      0.4  & 7.9  & 1.5    & 39 \\ \midrule
%			
		\end{tabular}
	\end{adjustbox}
\end{table}

\Table{\ref{tb:temps_encodage}} shows an average factor of 1,700 between real-time encoding and the \gls{hm} encoding time of \gls{hd} video content at \gls{qp}=32.
%In other words,the \gls{hm}-16.20 needs in average 1,800 seconds to encode 1 second of \gls{hd} video content.
This average factor is 16 time greater (27,000) with \gls{vtm}.
The \gls{vtm} main profile is therefore 16 times more complex in average compared to the \gls{hm} main profile for the encoding of \gls{hd} video content.
%As mentioned in Section~\ref{section:etat_art}, this encoding complexity increase is mainly caused by the \gls{mtt} partitioning in \gls{vtm}-5.0 which allows 4 additional partition modes compared to \gls{qt} partitioning in \gls{hm}-16.20.
This encoding complexity increase between \gls{hm}-16.20 and \gls{vtm}-5.0 is mainly caused by the the additional transform types and intra modes enabled in \gls{vtm}-5.0, and by the \gls{mtt} partitioning which allows 4 additional partition modes compared to \gls{qt} partitioning in \gls{hm}.

%In this work, the original sequences are down-sampled from \gls{uhd} to \gls{hd}, with the same frame-rate for both original and down-sampled sequences.
%The original \gls{uhd} sequences contain 4 times more \glspl{ctu} compared to down-sampled \gls{hd} sequence.
%\Table{\ref{tb:temps_encodage}} shows that the average ratio between the encoding time of \gls{uhd} and \gls{hd} sequences is 3.83 and 4.07 for \gls{hm}-16.20 and \gls{vtm}-5.0, respectively.
%This statistic confirms that the encoding time of \gls{hm}-16.20 and \gls{vtm}-5.0 is directly related with the number of \glspl{ctu}, since both encoders apply the same encoding process to each \gls{ctu}.

The results in \Table{\ref{tb:temps_encodage}} also highlight the impact of \gls{qp} value on encoding complexity.
Indeed, in order to avoid exhaustive \gls{rdo} process and to decrease encoding time for researchers experimentations, the reference encoders include early termination techniques.
In \gls{vtm} for instance, a dozen early termination techniques speed-up the encoding process by a factor 8~\cite{wieckowski_fast_2019} compared to exhaustive \gls{rdo} process.
The efficiency of these techniques vary with the encoding parameters, in particular with \gls{qp} value.
%This explains the impact of \gls{qp} values on encoding times observed in \Table{\ref{tb:temps_encodage}}.
%Indeed, encoding times carried out with lower \glspl{qp} are 44\% greater compared to encodings with higher \glspl{qp} for \gls{vtm}-5.0.
This explains why in \Table{\ref{tb:temps_encodage}}, for \gls{vtm}, encoding times carried out with lower \gls{qp} values are 44\% higher compared to encodings with higher \gls{qp} values.
The impact of \gls{qp} decreases to 11\% for \gls{hm}, due to \gls{qt} partitioning that offers less early termination opportunities compared to \gls{mtt} partitioning in \gls{vtm}.

Regardless the \gls{qp} value and resolution, encoding times standard deviations are around 20\% and 30\% of average encoding times for \gls{hm} and \gls{vtm}, respectively.
%The standard deviation is mainly induced by the diversity of frame-rates, \gls{si} and \gls{ti} values in the test sequences. 
These standard deviations are induced among others by the diversity of frame-rates among test sequences.
For instance, the sequence \emph{AerialCrowd2} with the lowest frame-rate (30~frames/sec) has the lowest encoding time in all columns of \Table{\ref{tb:temps_encodage}}.
The standard deviations are also a consequence of the diversity of spatial textures and movements among test sequences. 
%The frame-rate of sequences \emph{HorseJumping} and \emph{CrowdRun} is 50 frames/sec for both sequences.
Let us consider the sequences \emph{HorseJumping} and \emph{CrowdRun}, which have the same frame-rate of 50~frames/sec.
\Figure{\ref{fig:SI_TI}} shows that \emph{CrowdRun} has considerably higher \gls{si} and \gls{ti} compared to \emph{HorseJumping}.
%In all columns of \Table{\ref{tb:temps_encodage}}, \emph{CrowdRun} encoding time exceeds at least by a factor 1.4 \emph{HorseJumping} encoding time.
%This factor is due to the previously mentioned complexity reduction techniques included in the reference softwares, that are more efficient on sequences with low spatial textures and slow motion. \\
Given that previously mentioned early termination techniques are more efficient on sequences with lower spatial textures and slower motion, 
\emph{CrowdRun} encoding time exceeds at least by a factor 1.4 \emph{HorseJumping} encoding time, in all columns of \Table{\ref{tb:temps_encodage}}.

%Indeed, \Figure{\ref{fig:SI_TI}} shows that \emph{CrowdRun} has higher \gls{si} and \gls{ti} coordinates compared to \emph{HorseJumping}.

%As mentioned in Section~\ref{subsec:exp_setup}, the \gls{vtm}-5.0 features \gls{simd} optimizations that have been disabled for the present study. 
As mentioned in Section~\ref{subsec:exp_setup}, \gls{simd} optimizations have been disabled in the \gls{vtm} encoder in this work. Enabling the \gls{simd} optimizations speeds up the \gls{vtm} encoder by 1.9 times in average. This speed-up is very far from the 110,000 speed-up needed in average to achieve \gls{uhd} real time encoding. 
Complexity reduction of \gls{hevc} encoder has been widely investigated~\cite{tsang_mode_2019,shen_content-based_2019} in the past years, and recently more efforts have been dedicated to the \gls{vvc} encoder.
In fact, several techniques already proposed to reduce \gls{vtm} encoding complexity by predicting a reduced set of likely intra modes~\cite{zhang_fast_2017,ryu_machine_2018}, or by testing a reduced number of partition configurations~\cite{yang_low_2019,amestoy_random_2019}. We believe that the reduction of the \gls{vtm} encoding complexity will be an active research field over the next years.

\subsubsection{Encoding Complexity Repartition}\leavevmode\par\label{subsec:encoding_complexity}

%\begin{figure*}[!tbp]
 % \centering
%\begin{minipage}[b]{0.4\textwidth}
   % \includegraphics[width=\textwidth]{Subjectives/1hd}
  %\end{minipage}
%\vspace{0.6cm}
  %\begin{minipage}[b]{0.4\textwidth}
    %\includegraphics[width=\textwidth]{Subjectives/2hd}
  %\end{minipage}
%\begin{minipage}[b]{0.4\textwidth}
   % \includegraphics[width=\textwidth]{Subjectives/3uhd}
  %\end{minipage}
  %\begin{minipage}[b]{0.4\textwidth}
    %\includegraphics[width=\textwidth]{Subjectives/4uhd}
  %\end{minipage}
   %\caption{Example of subjective gains in termes of bit-rate with same perceived quality in two formats HD (top) and UHD (bottom).}
   %\label{Sub2}
%\end{figure*} 

\begin{figure*}[!t]		
	\begin{minipage}[b]{.45\linewidth}
		\begin{subfigure}[b]{\linewidth}
			\centerline{\includegraphics[width=0.7\linewidth]{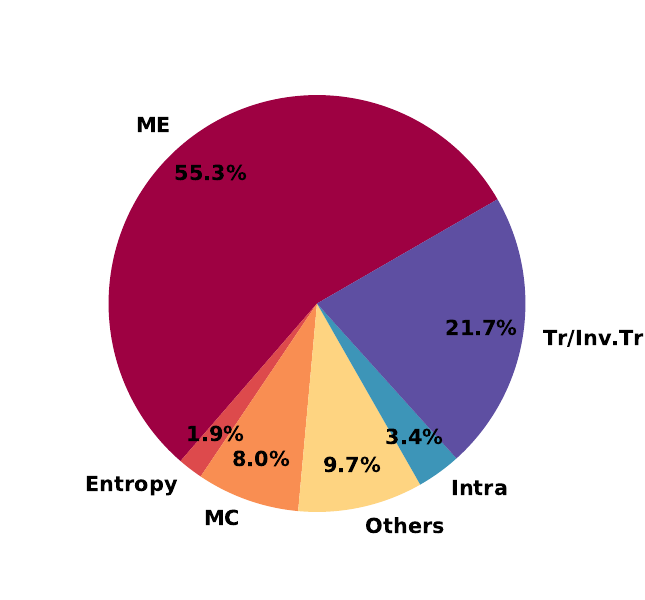}}
			\caption{\gls{hm}-16.20, QP=32 (8.6 Ks)}
		\end{subfigure}
	\end{minipage}\hfill
	\begin{minipage}[b]{.45\linewidth}	
		\begin{subfigure}[b]{\linewidth}
			\centerline{\includegraphics[width=0.7\linewidth]{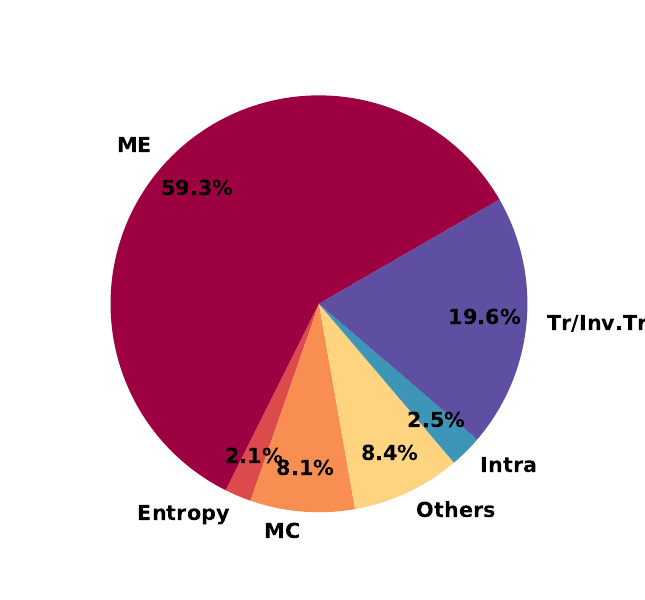}}
			\caption{\gls{hm}-16.20, QP=37 (7.6 Ks)}
		\end{subfigure}
	\end{minipage}
	\begin{minipage}[b]{.45\linewidth}
		\begin{subfigure}[b]{\linewidth}
			\centerline{\includegraphics[width=0.7\linewidth]{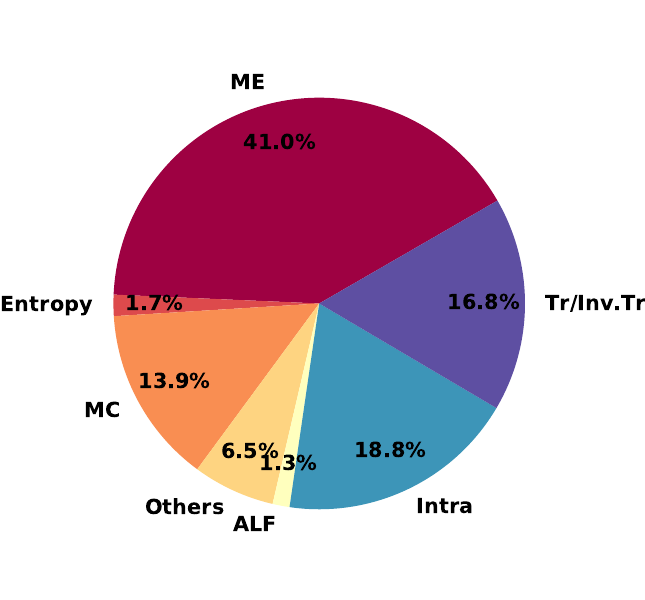}}
			\caption{\gls{vtm}-5.0, QP=32 (179 Ks)}
		\end{subfigure}
	\end{minipage}\hfill
	\begin{minipage}[b]{.45\linewidth}
		\begin{subfigure}[b]{\linewidth}
			\centerline{\includegraphics[width=0.7\linewidth]{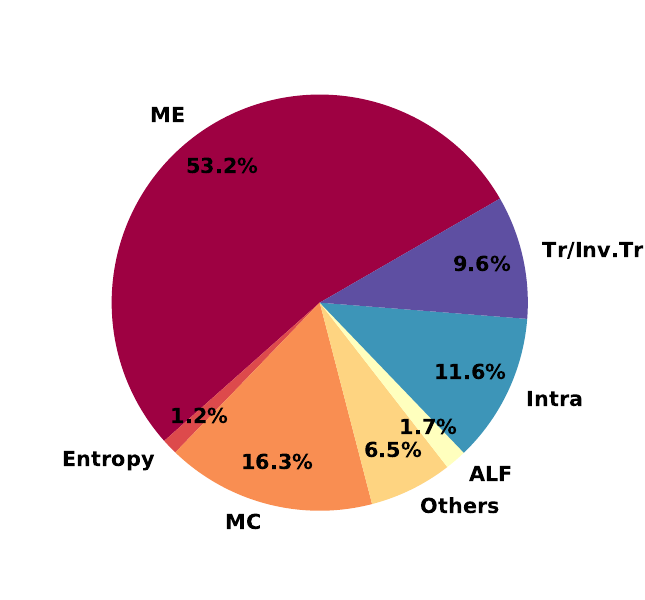}}
			\caption{\gls{vtm}-5.0, QP=37 (124 Ks)}
		\end{subfigure}
	\end{minipage}
	
	\caption{Encoding complexity repartition (in \%), for \gls{uhd} sequence CrowdRun, according to the reference software and \gls{qp} value.}
	\label{fig:profiling_encoder}
\end{figure*}

As mentioned in Section~\ref{subsec:exp_setup}, the encoding complexity repartition is obtained with Callgrind profiling tool.
The encoders of both \gls{hm}16.20 and \gls{vtm} have a significant portion of its complexity located in code cycles, since functions call each other in a recursive manner during the partitioning.
However, the graphical user interface for Callgrind data visualization, named KCachegrind, contains a cycle detection module that allows an easy profiling of encoding complexity.

\Figure{\ref{fig:profiling_encoder}} displays under the form of pie charts the encoding complexity repartition (in \%), for \gls{uhd} (4K) sequence CrowdRun, according to the reference software and \gls{qp} value.
The inter prediction stage is divided into 2 sub stages: \gls{me} and \gls{mc}.  
\gls{me} establishes the list of \gls{mv} predictor candidates for the processed \gls{cu} and then infers the best \gls{mv} predictor from this list.
\gls{mc} produces the prediction residuals by subtracting the inter predictions from the current original \glspl{cu}.
The \emph{Tr/Inv.Tr} stage sums the complexities induced by both Transform and Inverse Transform stages.
The stages representing less than 1\% of encoding complexity are gathered in the ``Other" category.

In \gls{ra} configurations, Intra prediction and Inter prediction are in competition for \glspl{pu} of P and B frames.
As expected, \Figure{\ref{fig:profiling_encoder}} shows that Inter prediction is heavily predominant compared to Intra prediction for both \gls{hm} and \gls{vtm}.
The percentage is even more unbalanced between Intra and Inter prediction when \gls{qp} value increases.
Indeed, Intra prediction percentage decreases from 3.4\% and 18.8\%, at \gls{qp}=32, to 2.5\% and 11.6\%, at \gls{qp}=37, while \gls{me} percentage raises from 55\% and 41\%, at \gls{qp}=32, to 59\% and 53\%, at \gls{qp}=37, for \gls{hm} and \gls{vtm}, respectively.
Therefore \Figure{\ref{fig:profiling_encoder}} highlights that the higher the \gls{qp} value, the more predominantly the encoder elects Inter prediction compared to Intra prediction.

In \Figure{\ref{fig:profiling_encoder}}, substantial differences are observable between \gls{hm} and \gls{vtm} complexity repartitions, especially for Intra Prediction and \gls{mc} stages.
The Intra Prediction percentage is increased by almost a factor 5 in \gls{vtm} compared to \gls{hm}. This increase is caused by the Intra prediction tools added in \gls{vtm} and mentioned in Section~\ref{Versatile Video Coding Description}, such as Multi-reference line prediction, Intra sub-partitioning and Matrix based intra prediction.
%\blue{The new Bi-directional optical flow and Affine motion model tools, also mentioned in Section~\ref{Versatile Video Coding Description}, are enabled during the \gls{mc} stage of \gls{vtm}-5.0 \red{(DMVR aussi, mais on n'en parle pas dans la section II).}} 
The Bi-directional optical flow, \gls{dmvr} and Affine motion model tools, also mentioned in Section~\ref{Versatile Video Coding Description}, have been introduced in the \gls{vtm} and are applied during the \gls{mc} stage.  
%These new tools partly explain the increase in \gls{mc} stage percentage in \gls{vtm}-5.0 compared to \gls{hm}-16.20, since they are responsible for 1.0\%, 1.4\% and 2.9\% of total \gls{vtm}-5.0 complexity, respectively.}
These new tools are responsible for 1.0\%, 1.4\% and 2.9\% of total \gls{vtm} complexity, respectively.
Therefore they partly explain the increase in \gls{mc} stage percentage in \gls{vtm} (around 15\%) compared to \gls{hm} (around 8\%).
%For this reason the \gls{mc} stage is imputable for a higher percentage in \gls{vtm}-5.0 (around 15\%) compared to \gls{hm}-16.20 (around 8\%).}
Finally, it is interesting to notice that the \gls{sao} and deblocking filters do not appear in \Figure{\ref{fig:profiling_encoder}} as their complexities are bellow 0.2\% for both \gls{hm} and \gls{vtm}.
However, \gls{alf} process must be taken into account in \gls{vtm}, since its complexity represents 1.3\% and 1.7\% of total encoding complexity at QP=32 and QP=37, respectively.

\subsection{Decoders Analysis}\label{sec:decoding_times_complexity}

\subsubsection{Decoding Time}\leavevmode\par\label{subsec:decoding_times}

\begin{table}[ht]
	\renewcommand{\arraystretch}{1.15}
	\centering
	\caption{Factor between decoding time and real time, for both \gls{vtm}-5.0 and \gls{hm}-16.20 in \gls{ra} configuration, according to the test sequence and \gls{qp} value.}
	\label{tb:temps_decodage}
	\begin{adjustbox}{max width=1\columnwidth}
		%		\begin{adjustbox}{max width=1\textwidth}
		
	\begin{tabular}{@{}l||cc|cc||cc|cc@{}}
		
		\multicolumn{1}{c}{}
		&\multicolumn{4}{c}{\textbf{\gls{hd}}}
		&\multicolumn{4}{c}{\textbf {\gls{uhd}} }  \\\midrule
		
		\textbf{Decoder } 
		&\multicolumn{2}{c|}{\textbf{\gls{hm}-16.20}}
		&\multicolumn{2}{c||}{\textbf{\gls{vtm}-5.0}}  
		&\multicolumn{2}{c|}{\textbf{\gls{hm}-16.20}}
		&\multicolumn{2}{c}{\textbf{\gls{vtm}-5.0}} \\\midrule
		
		\textbf{QP} & 
		\textbf{27}&  
		\textbf{32} &
		\textbf{27}&  
		\textbf{32} & 
		\textbf{32}&  
		\textbf{37} &
		\textbf{32}&  
		\textbf{37}    \\\midrule%\midrule
		
\emph{AerialCrowd2}& 2.8    & 2.3    & 8.8   & 7.8      & 8.4   & 7.3    & 27.9  & 25.7     \\
\emph{CatRobot1} &   4.1    & 3.6    & 13.1  & 11.1     & 13.1  & 12.1   & 44.4  & 35.5      \\
\emph{CrowdRun} &    5.4    & 4.4    & 15.4  & 13.1     & 14.2  & 12.1   & 49.0  & 39.6       \\
\emph{DaylightRoad}& 4.5    & 4.0    & 15.1  & 13.8     & 13.5  & 12.6   & 54.2  & 42.3       \\
\emph{Drums2} &      4.3    & 3.7    & 12.1  & 10.8     & 12.0  & 11.3   & 37.1  & 31.0       \\
\emph{HorseJumping}& 3.1    & 2.7    & 10.8  & 8.0      & 10.8  & 10.0   & 32.5  & 28.4      \\
\emph{Sedof}&        4.9    & 4.1    & 15.7  & 14.7     & 13.9  & 12.2   & 52.9  & 50.7     \\ \midrule
\textbf{\emph{Mean}} &      \textbf{4.1}    & \textbf{3.5}    & \textbf{13.0}  & \textbf{11.3}     & \textbf{12.3}  & \textbf{11.1}   & \textbf{41.4}  & \textbf{36.1}     \\ 
\textbf{\emph{Std Dev}} &     \textbf{0.9}    & \textbf{0.7}    & \textbf{2.5}   & \textbf{2.4 }     & \textbf{1.9}   & \textbf{1.7}    & \textbf{8.4}   & \textbf{8.0}   \\ \midrule
	
%	\begin{tabular}{@{}r||c|c||c|c@{}}
%		\multicolumn{1}{c}{}&\multicolumn{2}{c}{\textbf{\gls{hd}}}
%		&\multicolumn{2}{c}{\textbf {4K} }  \\\midrule
%		\textbf{Sequence name} & 
%		\textbf{\gls{hm}-16.20}&  
%		\textbf{\gls{vtm}-5.0} &
%		\textbf{\gls{hm}-16.20} &
%		\textbf{\gls{vtm}-5.0}   \\\midrule\midrule
%		
%		\emph{AerialCrowd2}& 4.0    & 8.8    & 14.1    & 30.0      \\
%		\emph{CatRobot1} &   6.9    & 13.1   & 25.5    & 48.3       \\
%		\emph{CrowdRun} &    8.0    & 14.3   & 27.3    & 49.2        \\
%		\emph{DaylightRoad}& 6.9    & 14.2   & 23.8    & 54.2        \\
%		\emph{Drums2} &      6.2    & 9.6    & 21.5    & 37.7        \\
%		\emph{HorseJumping}& 6.0    & 10.9   & 20.0    & 47.4       \\
%		\emph{Sedof}&        7.9    & 15.7   & 26.4    & 64.1            \\ \midrule\midrule
%		\emph{Mean } &       7.1    & 13.3   & 22.7    & 47.3      \\ \midrule
%		\emph{Std Dev} &     1.3    & 2.4    & 4.3     & 10.2    \\ \midrule
%		

		\end{tabular}
	\end{adjustbox}
\end{table}

The decoding process interprets the encoded symbols of a bitstream compliant with the standard specification.
It is therefore not burdened with the complex \gls{rdo} performed during encoding process that includes among others partitioning, intra mode decision or motion estimation.
%For this reason, comparatively to encoding time values presented in \Table{\ref{tb:temps_encodage}}, the \gls{hm}-16.20 decoding times shown in \Table{\ref{tb:temps_decodage}} are in average 500 times lower compared to encoding times, this ratio reaching 3000 for \gls{vtm}-5.0 decoding process.
For this reason, comparatively to encoding time values presented in \Table{\ref{tb:temps_encodage}}, the decoding times shown in \Table{\ref{tb:temps_decodage}} are in average 500 and 2000 times lower compared to encoding times for \gls{hm} and \gls{vtm}-5.0, respectively.

\begin{figure*}[t]

	\begin{minipage}[b]{.45\linewidth}
		\begin{subfigure}[b]{\linewidth}
			\centerline{\includegraphics[width=0.8\linewidth]{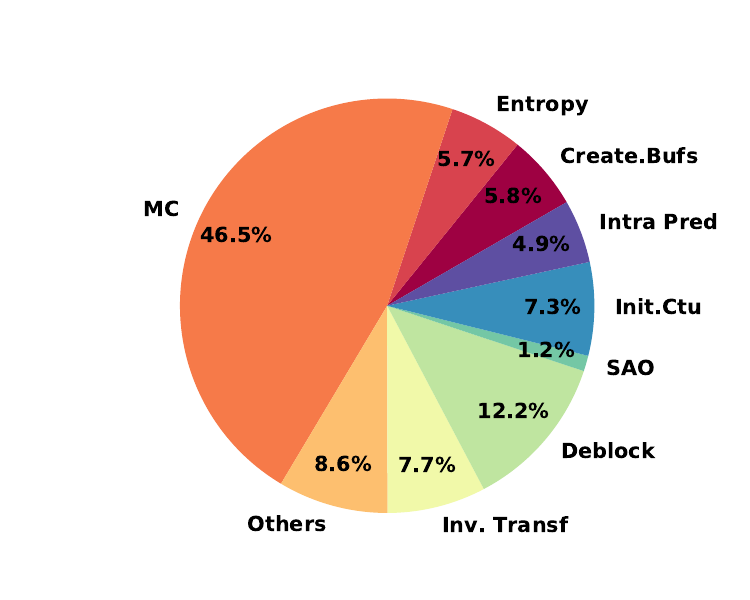}}
			\caption{\gls{hm}-16.20, QP=32 (12.3 s)}
		\end{subfigure}
	\end{minipage} \hfill
	\begin{minipage}[b]{.45\linewidth}
		\begin{subfigure}[b]{\linewidth}
			\centerline{\includegraphics[width=0.8\linewidth]{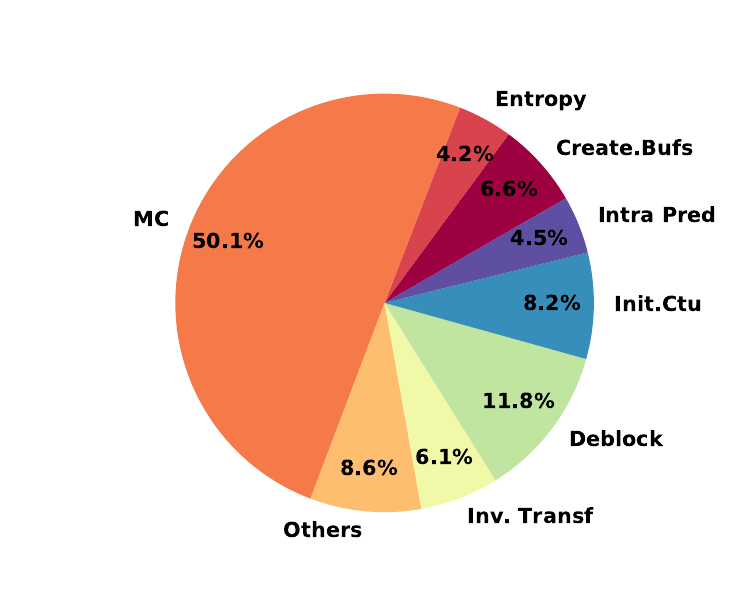}}
			\caption{\gls{hm}-16.20, QP=37 (11.1 s)}
		\end{subfigure}
	\end{minipage}
	
	\begin{minipage}[b]{.45\linewidth}
		\begin{subfigure}[b]{\linewidth}
			\centerline{\includegraphics[width=0.8\linewidth]{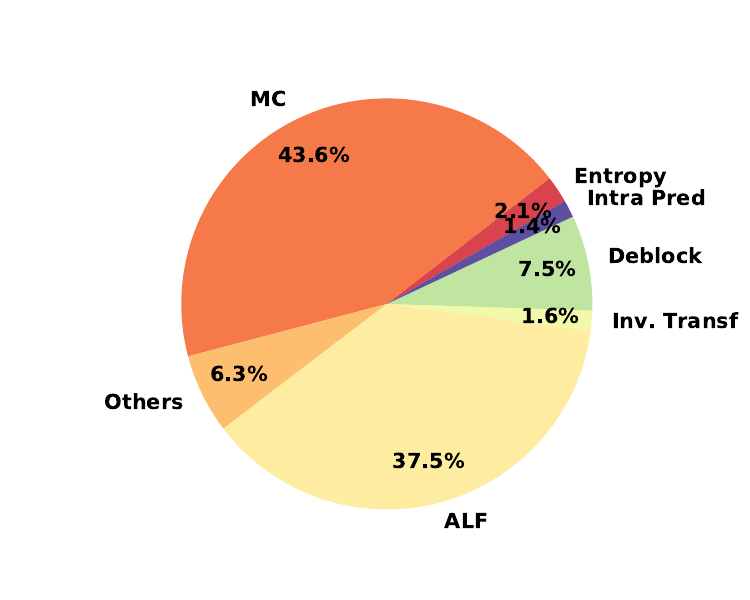}}
			\caption{\gls{vtm}-5.0, QP=32 (41.4 s)}
			\label{fig:decoupe_hevc}
		\end{subfigure}
	\end{minipage} \hfill
	\begin{minipage}[b]{.45\linewidth}
		\begin{subfigure}[b]{\linewidth}
			\centerline{\includegraphics[width=0.8\linewidth]{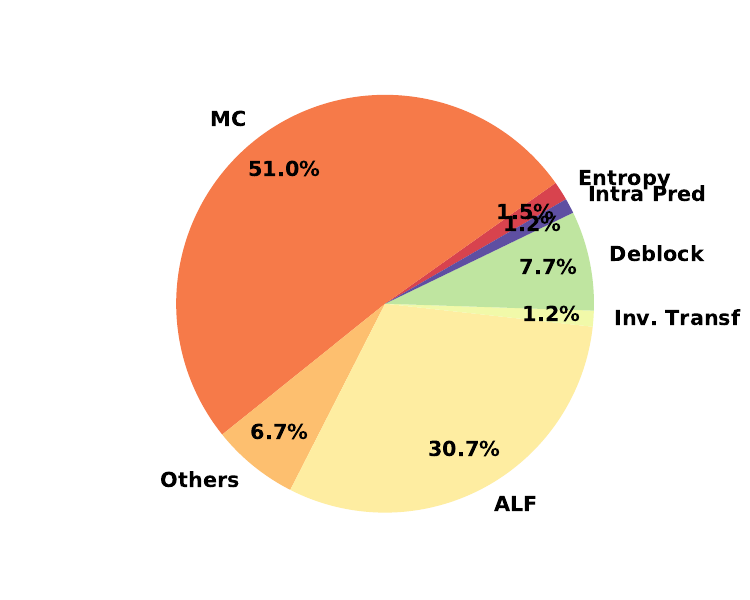}}
			\caption{\gls{vtm}-5.0, QP=37 (36.1 s)}
		\end{subfigure}
	\end{minipage}
	\caption{Decoding complexity repartition (in \%), averaged across \gls{uhd} test sequences, according to the reference software and \gls{qp} value.}
	\label{fig:profiling_decoder}
\end{figure*}
It is interesting to note that  the decoding time with \gls{vtm} is almost 3 times greater compared to the decoding time  with \gls{hm}.
This decoding complexity increase is partly imputable to the \gls{alf} filter introduced in \gls{vtm}, which complexity will be further discussed in Section~\ref{subsec:decoding_complexity}. 
Moreover, regardless of the resolution or test sequence, the decoding time decreases with the \gls{qp} value in \Table{\ref{tb:temps_decodage}} for both \gls{hm}-16.20 and \gls{vtm}-5.0.
Indeed, for a given sequence, a lower \gls{qp} value implies a larger bitstream outputted by the encoder, and therefore more symbols for the decoder to interpret especially at the level of by the \gls{cabac} engine.
%This explains why for all test sequences in \Table{\ref{tb:temps_decodage}}, the higher the \gls{qp} value, the lower the decoding time. 

%For \gls{uhd} video content, \Table{\ref{tb:temps_decodage}} shows that the decoding of 1 second takes 22.7 and 47.3 seconds in average with \gls{hm}-16.20 and \gls{vtm}-5.0, respectively.
%\red{Facteur 3.2 et 3.5 entre HD et \gls{uhd} . Du aux textures plus detaillees et decoupes plus fines sur HD ? Donc plus de temps par CTU sur HD ?}

As mentioned in Section~\ref{subsec:exp_setup}, \gls{simd} optimizations have been disabled at \gls{vtm} decoder side.
Enabling the \gls{simd} optimizations speeds up the \gls{vtm} decoder by 2.0 times in average for both \gls{hd} and \gls{uhd} sequences.
Therefore, by enabling \gls{simd} optimizations, the \gls{vtm} decoding time is only 1.5 times greater than \gls{hm} in average.
For \gls{hevc} standard, real-time decoding has been reached by adding hardware optimization, implementing assembly written functions and/or enabling high level parallelism. For instance, authors in~\cite{hamidouche_4k_2016} describe a real-time SHVC decoder based on the OpenHEVC open source decoder, and paper~\cite{bross_hevc_2013} presents a real-time decoder for \gls{hevc} standard relying on \gls{simd} optimizations and frame-level parallelism. 
In order to achieve real-time decoding for \gls{vvc} standard, similar efforts must be extended and improved in future years, since we have shown that decoding process is 3 times more complex for \gls{vvc} standard compared to \gls{hevc} standard.

\subsubsection{Decoding Complexity Repartition}\leavevmode\par\label{subsec:decoding_complexity}

%In order to obtain the encoding complexity repartition, the decoding process is ran with Callgrind~\cite{noauthor_callgrind_nodate}, a Valgrind profiling tool that records the call history of program functions as a call graph.
As for the encoder in Section~\ref{subsec:encoding_complexity}, the decoding complexity repartition is obtained by running the decoder with Callgrind~\cite{noauthor_callgrind_nodate}.
\Figure{\ref{fig:profiling_decoder}} displays the decoding complexity sharing (in \%), averaged across \gls{uhd} test sequences, according to the reference software and \gls{qp} value.
The Create Bufs. stage consists in the creation of global buffers at picture level, while Init Ctu creates internal buffers and sets initial values before encoding the \gls{ctu}.
The sum of these stages is greater than 13\% in \gls{hm}, but is negligible in \gls{vtm}.

The first noticeable fact among the presented pie charts is the predominance of \gls{mc}, ranging from 43.6\% to 51.0\% in both \gls{hm} and \gls{vtm}.
The predominance of Inter predicted frames in \gls{ra} configuration explains these numbers, and also explains the limited portion of Intra prediction in the pie charts.
In \gls{vtm}, an important share of \gls{mc} complexity lies in the \gls{dmvr}, that refines the motion vectors at decoder side, as mentioned in Section~\ref{Versatile Video Coding Description}.
In fact, the share of \gls{dmvr} in \gls{vtm} global decoding complexity is in average 23\% with \gls{qp}=32 and 32\% with \gls{qp}=37, which represents almost 60\% of \gls{mc} stage complexity.
However, the \gls{dmvr} is not significant burden as the previous percentages suggest for the decoding process. 
Indeed, when it is disabled, it is replaced to a large extent by usual \gls{mv} Inter prediction, which decoding complexity approaches \gls{dmvr} complexity. 
The contribution~\cite{JVET-O0013} confirms this behaviour, since it announces that \gls{vtm} decoding complexity is only 4\% lower when \gls{dmvr} is disabled compared to original \gls{vtm} in \gls{ra} configurations.
The deblocking filtering stage covers a consistent portion of the total decoding complexity: around 7.5\% for \gls{vtm} and around 12\% for \gls{hm}. In \gls{vtm}, the in-loop filtering percentage is heavily increased by additional \gls{alf} tool, which is responsible for around a third of total decoding complexity.
As mentioned in Section~\ref{Versatile Video Coding Description}, the \gls{alf} provides a significant improvement in encoding efficiency. It does not overload the \gls{vtm} encoding process by more than 2\% of total complexity (see \Figure{\ref{fig:profiling_encoder}}).
However, in counterpart for the aforementioned benefits, \gls{alf} represents a considerable burden for the decoding process. 
It is worth mentioning that \gls{alf} percentage decreases to less than 20\% when the \gls{simd} optimizations are enabled.

\section {Conclusion}
\label{sec:con}
In this paper, we have conducted a comprehensive investigation into the quality assessments and the associated computational complexities of the \gls{vvc} standard in comparison to the \gls{hevc} reference software. Our study focused on examining the bit-rate savings achieved between \gls{vtm} and the \gls{hm} software, as well as the distribution of encoding/decoding complexities across a typical classification of hybrid decoding tools. Our findings reveal that \gls{vtm}'s coding tools consistently deliver a substantial improvement in video quality compared to the \gls{hm} reference software across various video sequences used in our experiments. Utilizing quality metrics such as \gls{psnr}, \gls{ssim}, and \gls{vmaf}, we observed gains ranging from 31\% to 40\%, with the specific value depending on the chosen quality metric. Subjectively, \gls{vtm} significantly outperforms the \gls{hevc} reference software, particularly at lower bit-rates, achieving gains of up to 40\%. Furthermore, our analysis indicates that, for various sequences, the \gls{vvc} codec can provide equivalent perceived visual quality to \gls{hevc} while achieving bit-rate reductions of up to 50\%. However, it is important to note that we observed a substantial increase in computational complexity when comparing \gls{hm} to \gls{vtm}. This complexity increase is not uniform and varies depending on factors such as \gls{qp} values, frame rate, spatial texture, and motion. On the encoder side, our simultaneous profiling of both \gls{hm} and \gls{vtm} has identified additional intra modes and \gls{mtt} partitioning as the primary factors responsible for the complexity increase in \gls{vtm}. Additionally, on the decoder side, it was observed that \gls{alf} and \gls{dmvr} represent the main sources of additional computational load in \gls{vtm} compared to the \gls{hevc} decoding process.

\def\url#1{}
\bibliographystyle{IEEEbib}
\bibliography{Bibliography,Bibliography_quality_complexity}

\end{document}